\documentclass{article}%
\usepackage{amsfonts}
\usepackage{amsmath}%
\usepackage{amssymb}%
\usepackage{graphicx}
\newtheorem{theorem}{Theorem}

\newtheorem{corollary}[theorem]{Corollary}

\newtheorem{definition}[theorem]{Definition}

\newtheorem{lemma}[theorem]{Lemma}

\newtheorem{remark}[theorem]{Remark}

\newenvironment{proof}[1][Proof]{\textbf{#1.} }{\ \rule{0.5em}{0.5em}}
\begin{document}

\title{Separability preserving Dirac reductions of Poisson pencils on Riemannian manifolds}
\author{Maciej B\l aszak\thanks{Partially supported by Swedish Research Council Grant
No. 629-2002-6681} \thanks{Partially supported by KBN grant No. 5P03B 004 20}\\Institute of Physics, A. Mickiewicz University\\Umultowska 85, 61-614 Pozna\'{n}, Poland
\and Krzysztof Marciniak$^{\ast}$\thanks{On leave of absence from Department of
Physics, A. Mickiewicz University, Pozna\'{n}, Poland.}\\Department of Science and Technology \\Campus Norrk\"{o}ping, Link\"{o}ping University\\601-74 Norrk\"{o}ping, Sweden}
\maketitle

\begin{abstract}
Dirac deformation of Poisson operators of arbitrary rank is considered. The
question when Dirac reduction does not destroy linear Poisson pencils is
studied. A class of separability preserving Dirac reductions in the
corresponding quasi-bi-Hamiltonian systems of Benenti type is discussed. Two
examples of such reductions are given. This paper will appear in J. Phys. A: Math. Gen.

\end{abstract}


\section{Introduction}

Recently a new (quasi)-bi-Hamiltonian separability theory of Liouville
integrable finite dimensional systems was constructed \cite{morositondo}%
-\cite{zMaciejem} based on general properties of Poisson pencils on manifolds.
A natural further step within this theory is to investigate admissible
integrable and separable reductions of these integrable / separable systems
onto appropriate submanifolds. The most natural approach seems to be the one
based on the Dirac theory of constrained dynamics \cite{Dirac},\cite{Dirac1}.
The presented paper contains only some special cases of such reductions, but
even in these cases the problem is far from being trivial. The difficulties we
met during our research inclined us to reconsider the Dirac formalism from the
point of view of Poisson bivectors rather then from the point of view of
constrained dynamics.

The paper is organized as follows. In this introductory part we remind basic
concepts of Poisson geometry and of (quasi-)bi-Hamiltonian systems. In Section
2 we formulate the theory of Dirac reductions of Poisson brackets in terms of
Poisson bivectors. Our construction is more general than usually met in
literature since we consider reduction procedure on the
whole\emph{\ foliation} of submanifolds. Here we also explain that our
approach indicates that Dirac classification of constraints onto those of
"first" and of "second class" requires further discussion as our further
results show. In Section 3 we review the recent results on separability theory
of quasi-bi-Hamiltonian systems of Benenti type. In section 4 we perform Dirac
reduction of Poisson pencil and corresponding quasi-bi-Hamiltonian chain of
Benenti type onto a particularly chosen submanifold. Chosen constraints
preserve the Liouville integrability as well as the coordinates of separation
of considered system. The main obstacle of such a choice is that these
constraints are nonexpressible in natural (original) coordinates. Hence, in
Section 5 we modify constraints introduced in the previous section and obtain
an equivalent reduction, that is expressible directly in original coordinates.
Finally, in Section 6 we illustrate the results by two nontrivial examples of
constrained separable dynamics.

Let us first remind few basic facts from Poisson geometry. Given a manifold
$\mathcal{M}$, a \emph{Poisson operator} $\pi$ on $\mathcal{M}$ is a mapping
$\pi:T^{\ast}\mathcal{M}\rightarrow T\mathcal{M}$ that is fibre-preserving
(i.e. $\pi|_{T_{x}^{\ast}\mathcal{M}}:T_{x}^{\ast}\mathcal{M}\rightarrow
T_{x}\mathcal{M}$ for any $x\in\mathcal{M}$) and such that the induced bracket
on the space $C^{\infty}(\mathcal{M})$ of all smooth real-valued functions on
$\mathcal{M}$
\begin{equation}
\left\{  .,.\right\}  _{\pi}:C^{\infty}(\mathcal{M})\times C^{\infty
}(\mathcal{M})\rightarrow C^{\infty}(\mathcal{M})\text{ \ , \ }\left\{
F,G\right\}  _{\pi}\overset{\mathrm{def}}{=}\left\langle dF,\pi
\,dG\right\rangle \label{bracket}%
\end{equation}
(where $\left\langle .,.\right\rangle $ is the dual map between $T\mathcal{M}$
and $T^{\ast}\mathcal{M}$) is skew-symmetric and satisfies Jacobi identity
(the bracket (\ref{bracket}) always satisfies the Leibniz rule $\left\{
F,GH\right\}  _{\pi}=G\left\{  F,H\right\}  _{\pi}+H\left\{  F,G\right\}
_{\pi}$). Throughout the whole paper the symbol $d$ will denote the operator
of exterior derivative. The operator $\pi$ can always be interpreted as a
bivector, $\pi\in\Lambda^{2}(\mathcal{M})$ and in a given coordinate system
$(x^{1},\ldots,x^{m})$ on $\mathcal{M}$ we have
\[
\pi=\sum\limits_{i<j}^{m}\pi^{ij}\frac{\partial}{\partial x_{i}}\wedge
\frac{\partial}{\partial x_{j}}.
\]
A function $C:\mathcal{M}\rightarrow\mathbb{R}$ is called \emph{Casimir
function} of the Poisson operator $\pi$ if for arbitrary function
$F:\mathcal{M}\rightarrow\mathbb{R}$ we have $\left\{  F,C\right\}  _{\pi}=0$
(or, equivalently, if $\pi dC=0$). A linear combination $\pi_{\xi}=\pi_{1}%
-\xi\pi_{0}$ ($\xi\in\mathbb{R}$) of two Poisson operators $\pi_{0}$ and
$\pi_{1}$ is called \emph{Poisson pencil} if the operator $\pi_{\xi}$ is
Poisson for any value of the parameter $\xi$. In this case we say that
$\pi_{0}$ and $\pi_{1}$ are \emph{compatible}$.$ Given a Poisson pencil
$\pi_{\xi}=\pi_{1}-\xi\pi_{0}$ we can often construct a sequence of \ vector
fields $Y_{i}$ on $\mathcal{M}$ that have a twofold Hamiltonian form (a so
called $\emph{bi-Hamiltonian}$ $\emph{chain}$)
\begin{equation}
Y_{i}=\pi_{1}dh_{i}=\pi_{0}dh_{i+1}\label{chain}%
\end{equation}
where $h_{i}:\mathcal{M}\rightarrow\mathbb{R}$ are called Hamiltonians of the
chain (\ref{chain}) and where $i\ $\ is some discrete index. This sequence of
vector fields may or may not truncate (depending on existence of Casimir
functions). In the case when the Poisson pencil $\pi_{\xi}$ is degenerated but
projectable into a symplectic leaf $\mathcal{N}$ (of dimension $2n$) of
$\pi_{0}$ the bi-Hamiltonian chain (\ref{chain}) on $\mathcal{M}$ turns into a
so called quasi-bi-Hamiltonian chain on $\mathcal{N}$ of the form%
\begin{equation}
\theta_{1}dH_{i}=\theta_{0}dH_{i+1}+\sum_{\text{ }\overset{j=1}{j\neq i+1}%
}^{n}\alpha_{ij}\theta_{0}dH_{j},\,\text{\ \ \ }i=1,\ldots,n\text{, \ }%
H_{n+1}\equiv0,\label{quasichain}%
\end{equation}
where $\theta_{i}$ are projections of $\pi_{i}$ onto $\mathcal{N}$, the
functions $H_{j}$ are restrictions of $h_{j}$ to $\mathcal{N}$: $H_{j}%
=h_{j}|_{\mathcal{N}}$, and $\alpha_{ij}$ are some multipliers (real
functions). And vice versa: having a quasi-bi-Hamiltonian chain
(\ref{quasichain}) on the manifold $\mathcal{N}$ one can lift it to a
bi-Hamiltonian chain (\ref{chain}) on the extended manifold $\mathcal{M}$.
(Quasi-)bi-Hamiltonian chains (called also (quasi-)bi-Hamiltonian systems)
possess very interesting differential-algebraic properties and are one of key
notions in the theory of integrable systems, due to the fact that in many
cases the systems (\ref{chain}) and (\ref{quasichain}) are Liouville
integrable \cite{Magri}. Recently, much effort has been spent in order to
exploit the procedure of solving these systems by the method of separation of
variables \cite{1}-\cite{zMaciejem}. In this article we will mainly work with
the quasi-bi-Hamiltonian chains (\ref{quasichain}) rather than bi-Hamiltonian
ones, since the pencil $\theta_{\xi}=\theta_{1}-\xi\theta_{0}$ is always
non-degenerated$.$

\section{Dirac reduction of Poisson bivectors}

We begin by considering the Dirac reduction procedure in a more general
setting that is usually met in literature. Let $\pi$ be a Poisson bivector, in
general degenerated on some manifold $\mathcal{M}.$ Let $\mathcal{S=F}_{0} $
be a submanifold in a foliation $\mathcal{F}$ of the manifold $\mathcal{M}$
defined by $m$ functionally independent functions (constraints) $\varphi
_{i}:\mathcal{M}\rightarrow\mathbb{R},\;i=1,\ldots,m.$%

\[
\mathcal{F}_{s}=\left\{  x\in M:\varphi_{i}(x)=s_{i},\,i=1,\ldots,m\right\}
\]
Thus, $\mathcal{S}$ is a submanifold of codimension $m$ in $\mathcal{M}.$
Moreover, let $Z_{i},\,i=1,...,m$ be some vector fields transversal to
$\mathcal{F}_{s}$, spanning a regular distribution $\mathcal{Z}$ in
$\mathcal{M}$ of constant dimension $m$ (that is a smooth collection of
$m$-dimensional subspaces $\mathcal{Z}_{x}\subset$ $T_{x}\mathcal{M}$ at every
point $x$ in $\mathcal{M}$). The word 'transversal' means here that no vector
field $Z_{i}$ is at any point tangent to the submanifold $\mathcal{F}_{s}$
passing through this point. Hence, the tangent bundle $T\mathcal{M}$ splits
into a direct sum
\[
T\mathcal{M}=T\mathcal{F}\oplus\mathcal{Z}%
\]
(which means that at any point $x$ in $\mathcal{M}$ we have $T_{x}%
\mathcal{M}=T_{x}\mathcal{F}_{s}\oplus\mathcal{Z}_{x}$ with $s$ such that
$x\in\mathcal{F}_{s}$ ) and so does its dual
\[
T^{\ast}\mathcal{M}=T^{\ast}\mathcal{F}\oplus\mathcal{Z}^{\ast},
\]
where $T^{\ast}\mathcal{F}$ is the annihilator of $\mathcal{Z}$ and
$\mathcal{Z}^{\ast}$ is the annihilator of $T\mathcal{F}$. That means that if
$\alpha$ is a one form in $T^{\ast}\mathcal{F}$ then $\alpha(Z_{i})=0$ for all
$i=1,\ldots,m$ and if $\beta$ is a one-form in $\mathcal{Z}^{\ast} $ then
$\beta$ vanishes on all vector fields tangent to $\mathcal{F}_{s}$. Moreover,
we assume that the vector fields $Z_{i}$ which span $\mathcal{Z}$ are chosen
in such a way that $d\varphi_{i},\;i=1,...,m$ \ is a basis in $\mathcal{Z}%
^{\ast}$ that is dual to the basis $Z_{i}$ of the distribution $\mathcal{Z}$,
\begin{equation}
\left\langle d\varphi_{i},Z_{j}\right\rangle =Z_{j}(\varphi_{i})=\delta
_{ij},\label{2.3}%
\end{equation}
(this is no restriction since for any distribution $\mathcal{Z}$ transversal
to $\mathcal{F}_{s}$ we can choose its basis so that (\ref{2.3}) is
satisfied). Finally, let us define $m$ vector fields $X_{i}$ on $\mathcal{M}$
and $m^{2}$ functions $\varphi_{ij}:\mathcal{M}\rightarrow\mathbb{R}$ on
$\mathcal{M}$ through
\begin{equation}
X_{i}=\pi(d\varphi_{i}),\;\;\ \ \;\varphi_{ij}=\{\varphi_{i},\varphi
_{j}\}_{\pi}=\,\left\langle d\varphi_{i},\pi\,d\varphi_{j}\right\rangle
=X_{j}(\varphi_{i}).\label{2.4}%
\end{equation}
The functions $\varphi_{ij}$ define an $m$-dimensional skew-symmetric matrix
$\varphi=\left(  \varphi_{ij}\right)  ,$ $i,j=1,\ldots m$. It can be easily
shown that
\begin{equation}
\lbrack X_{j},X_{i}]=X_{\{\varphi_{i},\varphi_{j}\}_{\pi}}=\pi\,\,d\{\varphi
_{i},\varphi_{j}\}_{\pi}=\pi\,d\varphi_{ij},\label{2.5}%
\end{equation}
where $\{.,.\}_{\pi}$ is a Poisson bracket defined by our Poisson bivector
$\pi$ and $[X,Y]=L_{X}Y=X(Y)-Y(X)$ is the Lie bracket (commutator) of the
vector fields $X,Y$.

A very special choice of our transversal vector fields $Z_{i}$ originates by
taking linear combinations of fields $X_{j}$ with coefficients \emph{being the
entries of the matrix} $\varphi^{-1}$.
\begin{equation}
Z_{i}=\sum_{j=1}^{m}(\varphi^{-1})_{ji}X_{j}\ \ \ \ \ \ i=1,...,m.\label{2.8}%
\end{equation}
Since the constraint functions $\varphi_{i}$ are functionally independent, the
vector fields $Z_{i}$ in (\ref{2.8}) will indeed be transversal to the
foliation $\mathcal{F}$. Moreover, they will automatically satisfy the
orthogonality condition (\ref{2.3}) as $Z_{i}(\varphi_{j})=\sum_{k=1}%
^{m}(\varphi^{-1})_{ki}X_{k}(\varphi_{j})$ $=\sum_{k=1}^{m}(\varphi^{-1}%
)_{ki}\varphi_{jk}=\delta_{ij}$. $\ $

Let us now consider the following \emph{deformation} (modification) of the
bivector $\pi$:
\begin{equation}
\pi_{D}=\pi-\tfrac{1}{2}\sum_{i=1}^{m}X_{i}\wedge Z_{i},\label{2.10}%
\end{equation}
where $\wedge$ denotes the wedge product in the algebra of multivectors. This
new bi-vector $\pi_{D}$ can be properly restricted to $\mathcal{F}_{s}$ for
any $s\in\mathbb{R}^{m\text{ }}$(and thus also to $\mathcal{S=F}_{0}$), since
the image of $\pi_{D}$ considered on a given leaf $\mathcal{F}_{s}$ of the
foliation $\mathcal{F}$ lies in $T\mathcal{F}_{s}$ for all $s$. This is the
content of the following theorem.

\begin{theorem}
\label{obciecie} Suppose $x\in\mathcal{F}_{s}$. Then for any $\alpha\in
T_{x}^{\ast}\mathcal{M}$ the vector $\pi_{D}(\alpha)$ is tangent to
$\mathcal{F}_{s}$ i.e. $\pi_{D}(T_{x}^{\ast}\mathcal{M})\subset T_{x}%
\mathcal{F}_{s}.$
\end{theorem}

\begin{proof}
We will show, that $\pi_{D}(d\varphi_{k})=0,\;k=1,...,m$, since it means that
the constraints $\varphi_{i},$ $i=1,...,m$ are Casimirs of $\pi_{D}$ (and so
are then $\varphi_{i}-s_{i}$) which obviously implies the thesis of the
theorem. Using the definition (\ref{2.10}) of $\pi_{D}$, the obvious fact that
$(X_{i}\wedge Z_{i})d\varphi_{k}=(X_{i}\otimes Z_{i})d\varphi_{k}%
-(Z_{i}\otimes X_{i})d\varphi_{k}=\left\langle d\varphi_{k},Z_{i}\right\rangle
X_{i}-\left\langle d\varphi_{k},X_{i}\right\rangle Z_{i}=Z_{i}(\varphi
_{k})X_{i}-X_{i}(\varphi_{k})Z_{i}$ we have
\begin{align*}
\pi_{D}(d\varphi_{k})  & =\pi(d\varphi_{k})-\tfrac{1}{2}\sum_{i}Z_{i}%
(\varphi_{k})X_{i}+\tfrac{1}{2}\sum_{i}X_{i}(\varphi_{k})Z_{i}\\
& =X_{k}-\tfrac{1}{2}\sum_{i}\delta_{ik}X_{i}+\tfrac{1}{2}\sum_{i,j}%
\varphi_{ki}(\varphi^{-1})_{ji}X_{j}\\
& =X_{k}-\tfrac{1}{2}X_{k}-\tfrac{1}{2}\sum_{i}\delta_{ki}X_{i}=0
\end{align*}
due to the fact that $\varphi_{ij}=-\varphi_{ji}.$
\end{proof}

\begin{theorem}
The bivector $\pi_{D}$ in (\ref{2.10}) with $Z_{i}$ as in (\ref{2.8})
satisfies the Jacobi identity.
\end{theorem}

\begin{proof}
It is easy to check that our operator $\pi_{D}$ defines the following bracket
on $\mathcal{M}$
\begin{equation}
\{F,G\}_{\pi_{D}}=\{F,G\}_{\pi}-\sum_{i,j=1}^{m}\{F,\varphi_{i}\}_{\pi
}(\varphi^{-1})_{ij}\{\varphi_{j},G\}_{\pi},\label{2.17}%
\end{equation}
(where $F,G:\mathcal{M}\rightarrow\mathbb{R}$ are two arbitrary functions on
$\mathcal{M}$) which is just the well known \emph{Dirac deformation}
\cite{Dirac} of the bracket $\{.,.\}_{\pi}$ associated with $\pi$, and as was
shown by Dirac \cite{Dirac1} it satisfies the Jacobi identity.
\end{proof}

\begin{remark}
\label{Diracwlas}If $C:\mathcal{M}\rightarrow\mathbb{R}$ is a Casimir function
of $\pi$, then it is also a Casimir function of $\pi_{D}$, since in this case
(\ref{2.17}) yields
\[
\{F,C\}_{\pi_{D}}=\{F,C\}_{\pi}-\sum_{i,j=1}^{m}\{F,\varphi_{i}\}_{\pi
}(\varphi^{-1})_{ij}\{\varphi_{j},C\}_{\pi}=0-0=0.
\]
We also know from Theorem \ref{obciecie} that the constraints $\varphi_{i}$
are Casimirs of the deformed operator $\pi_{D}$. Thus, we can informally state
that Dirac deformation preserves all the old Casimir functions and introduces
new Casimirs $\varphi_{i}$.
\end{remark}

It is now possible to restrict our Poisson operator $\pi_{D}$ (or our Poisson
bracket $\{.,.\}_{\pi_{D}}$) to a Poisson operator (bracket) on the
submanifold $\mathcal{S}$ (i.e. a symplectic leaf $\varphi_{1}=...=\varphi
_{m}=0$ of $\pi_{D}$) in a standard way (reduction to a symplectic leaf).
Namely, for arbitrary functions $f,g:\mathcal{S}\rightarrow R$ one defines the
\emph{reduced Dirac bracket} $\{f,g\}_{R}$ on $\mathcal{S}$ (with the
corresponding Poisson operator $\pi_{R}$)  as the restriction of the bracket
of arbitrary prolongations of $f$ and $g$ to $\mathcal{M}$, i.e.
\[
\{f,g\}_{R}=\{F,G\}_{D}|_{\mathcal{S}}%
\]
where $F,G:\mathcal{M}\rightarrow\mathbb{R}$ are two functions such that
$f=F|_{\mathcal{S}}$ and $g=G|_{\mathcal{S}}$. Of course, an identical
construction can be induced on any leaf $\mathcal{F}_{s}$ from the foliation
$\mathcal{F}$. There arises, of course, a question how 'robust' this
construction is, i.e. to what extent the obtained deformation $\pi_{D}$ and
the reduction $\pi_{R}$ are independent of the choice of particular functions
$\varphi_{i}$ that define our submanifold $S.$ This issue will be partially
addressed in next sections.

The results of this section as well as the results presented in Sections 4 and
5 suggest that the concept of the classification of constraints as being
either of \textquotedblright first-class\textquotedblright\ or of
\textquotedblright second-class\textquotedblright, proposed by Dirac, should
be reexamined when one looks on the problem from the point of view of Poisson
geometry. First of all it is clear that the procedure called \emph{Dirac
reduction} has two different levels. The first level we call \emph{Dirac
deformation }as we deform a Poisson bivector $\pi$ from manifold $\mathcal{M}$
to another Poisson bivector $\pi_{D}$ on the same manifold $\mathcal{M}$. A
sufficient condition for the existence of $\pi_{D\text{ }}$ for a given set of
constraints $\varphi_{i},$ $i=1,...,m$ is a nondegeneracy of the Gram matrix
$\varphi.$ From the construction all constraints $\varphi_{i}$ are Casimirs of
$\pi_{D}.$The second level of the construction we call \emph{Dirac restriction
}as we restrict a Poisson bivector $\pi_{D}$ to its symplectic leaf
$\mathcal{F}_{s}.$ In the original Dirac construction it was a particular one,
namely $\mathcal{F}_{0}=\mathcal{S}$. A Poisson bivector on $\mathcal{F}_{s}$
is denoted by $\pi_{R}.$ For the existence of the second step additional
restrictions on $\varphi_{i}$ have to be imposed. Actually, $\pi_{R}$ has to
be nonsingular. In a standard classification it means that we have to exclude
constraints of first-class. Let us remind that a constraint $\varphi_{k}$ is
of \emph{first class} if its Poisson bracket with all the remaining constants
$\varphi_{i}$ vanishes on $\mathcal{S}$, that is if
\begin{equation}
\{\varphi_{k},\varphi_{i}\}_{\pi}|_{\mathcal{S}}%
=0,\;\;\;\;\;i=1,...,m.\label{first}%
\end{equation}
Otherwise $\varphi_{k}$ is of \emph{second-class}. In general, first-class
constraints make $\pi_{R}$ singular so that the Dirac reduction procedure can
not be performed. Nevertheless, the condition (\ref{first}) seems to be too
strong. In sections 4 and 5 we demonstrate situations where constraints are of
first-class but the singularity is 'removable' and so the Poisson bivector
$\pi_{R}$ is well defined. It suggests that the classification of constraints
given by Dirac should be reformulated in the context of Poisson pencils and
Poisson geometry.

\section{Separable quasi-bi-Hamiltonian chains of Benenti type}

In the following section we briefly remind basic facts about separable
Hamiltonian systems on Riemannian manifolds, which form a special class of
quasi-bi-Hamiltonian chains \cite{2},\cite{ibort}, known also as the so called
Benenti systems \cite{b1},\cite{b2}. Let $(Q,g)$ be a Riemannian manifold with
covariant metric tensor $g=(g_{ij})$ and with the inverse (contravariant
metric tensor) $g^{-1}=G=(G^{ij})$. Let $(q^{1},...,q^{n})$ be some coordinate
system on $Q$ and let \ $(q^{1},...,q^{n},p_{1},...,p_{n})$ be the
corresponding canonical coordinates on the phase space $\mathcal{N}=T^{\ast}Q$
with the associated Poisson tensor
\begin{equation}
\theta_{0}=\left(
\begin{array}
[c]{cc}%
0_{n} & I_{n}\\
-I_{n} & 0_{n}%
\end{array}
\right)  \label{3.1}%
\end{equation}
where $I_{n}$ is $n\times n$ unit matrix and $0_{n}$ is the $n\times n$ matrix
with all entries equal to zero. Let us consider the Hamiltonian $E:\mathcal{N}%
\rightarrow\mathbb{R}$ for the geodesic flow on $Q$:
\begin{equation}
E=E(q,p)=\sum_{i,j=1}^{n}G^{ij}(q)p_{i}p_{j}.\label{3.2}%
\end{equation}
As it is known, a $(1,1)$-type tensor $B=(B_{j}^{i})$ (or a $(2,0)$-type
tensor $B=(B^{ij}))$ is called a Killing tensor with respect to $g$ if
$\left\{  \,%
%TCIMACRO{\tsum }%
%BeginExpansion
{\textstyle\sum}
%EndExpansion
(BG)^{ij}p_{i}p_{j}\,,\,E\right\}  _{\theta_{0}}=0$ \ \ (or $\left\{  \,%
%TCIMACRO{\tsum }%
%BeginExpansion
{\textstyle\sum}
%EndExpansion
(B)^{ij}p_{i}p_{j}\,,\,E\right\}  _{\theta_{0}}=0$). An important
generalization of this notion is formulated in the following definition.

\begin{definition}
\label{conformalKilling}Let $L=(L_{j}^{i})$ be a second order mixed type (i.e.
$(1,1)$-) tensor on $Q$ and let $\overline{L}:\mathcal{N\rightarrow}%
\mathbb{R}$ be a function on $\mathcal{N}$ defined as $\overline{L}=\frac
{1}{2}%
%TCIMACRO{\dsum \limits_{i,j=1}^{n}}%
%BeginExpansion
{\displaystyle\sum\limits_{i,j=1}^{n}}
%EndExpansion
(LG)^{ij}p_{i}p_{j}$ , where $LG$ is a $(1,1)$ tensor with components
$(LG)^{ij}=%
%TCIMACRO{\dsum \limits_{i,j=1}^{n}}%
%BeginExpansion
{\displaystyle\sum\limits_{i,j=1}^{n}}
%EndExpansion
L_{k}^{i}g^{kj}$. If
\[
\{\overline{L},E\}_{\theta_{0}}=\alpha E,\ \ \ \ \mathrm{\ where}%
\ \ \ \ \ \ \alpha=\sum_{i,j=1}^{n}G^{ij}\frac{\partial f}{\partial q^{i}%
}p_{j},\ \ \ \ f=Tr(L),
\]
then $L$ is called a \emph{special conformal Killing tensor} with the
associated potential $f=Tr(L)$ \cite{ibort}.
\end{definition}

The importance of this notion lies in the fact that on manifolds with tensor
$L$ the geodesic flows are separable. There exist, in this case, $n$ constants
of motion, quadratic in momenta, of the form
\begin{equation}
E_{r}=\sum_{i,j=1}^{n}A_{r}^{ij}p_{i}p_{j}=\sum_{i,j=1}^{n}(K_{r}G)^{ij}%
p_{i}p_{j},\ \ \ \ \ \ \ \ \ \ r=1,...,n,\label{3.4}%
\end{equation}%
\[
(K_{r}G)^{ij}=%
%TCIMACRO{\dsum \limits_{k=1}^{n}}%
%BeginExpansion
{\displaystyle\sum\limits_{k=1}^{n}}
%EndExpansion
(K_{r})_{k}^{i}G^{kj}%
\]
where $A_{r}$ and $K_{r}$ are Killing tensors of type $(2,0)$ and $(1,1)$,
respectively. Moreover, all the Killing tensors $K_{r}$ are given by the
following 'cofactor' formula
\begin{equation}
cof(\xi I-L)=\sum_{i=1}^{n-1}K_{n-i}\xi^{i},\label{3.4a}%
\end{equation}
where $cof(A)$ stands for the matrix of cofactors, so that $cof(A)A=(\det
A)I.$ Notice that $K_{1}=I,$ hence $A_{1}=G$ and $E_{1}\equiv E.$ Since the
tensors $K_{r}$ are Killing, with a common set of eigenfunctions, the
functions $E_{r}$ satisfy $\{E_{s},E_{r}\}_{\theta_{0}}=0$ and thus they
constitute a system of $n$ constants of motion in involution with respect to
the Poisson structure $\theta_{0}$. So, for a given metric tensor $g$, the
existence of a special conformal Killing tensor $L$ is a sufficient condition
for the geodesic flow on $\mathcal{N}$ to be a Liouville integrable
Hamiltonian system.

The special conformal Killing tensor $L$ can be lifted from $Q$ to a
$(1,1)$-type tensor on $\mathcal{N}=T^{\ast}Q$ where it takes the form
\begin{equation}
N=\left(
\begin{array}
[c]{cc}%
L & 0_{n}\\
F & L^{T}%
\end{array}
\right)  ,\;\;\;\;F_{j}^{i}=\frac{\partial}{\partial q^{i}}(Lp)_{j}%
-\frac{\partial}{\partial q^{j}}(p^{T}L)_{i}.\label{3.14}%
\end{equation}
The lifted $(1,1)$ tensor $N$ is called a \emph{recursion operator}%
.\emph{\ }An important property of $N$ is that when it acts on the canonical
Poisson tensor $\theta_{0}$ it produces another Poisson tensor
\[
\theta_{1}=N\theta_{0}=\left(
\begin{array}
[c]{cc}%
0 & L\\
-L^{T} & F
\end{array}
\right)  ,
\]
compatible with the canonical one (actually $\theta_{0}$ is compatible with
$N^{k}\theta_{0}$ for any integer $k$).

It is now possible to show that the geodesic Hamiltonians $E_{r}$ satisfy on
$\mathcal{N}=T^{\ast}Q$ the set of relations \emph{\ }
\begin{equation}
\theta_{0}dE_{r+1}=\theta_{1}dE_{r}+\rho_{r}\theta_{0}dE_{1},\;\;\;\;E_{n+1}%
=0,\;\;\;\;\;\;r=1,...,n,\label{3.17}%
\end{equation}
where the functions $\rho_{r}(q)$ are coefficients of the characteristic
polynomial of $L$ (i.e. minimal polynomial of $N$), which is a special case of
the quasi-bi-Hamiltonian chain (\ref{quasichain})\emph{\ }\cite{morositondo}%
$.$

It turns out that with the tensor $L$ we can (generically) associate a
coordinate system on $\mathcal{N}$ in which the flows associated with all the
functions $E_{r}$ separate. Namely, let $(\lambda^{1}(q),...,\lambda^{n}(q))$
be $n$ distinct, functionally independent eigenvalues of $L$, i.e. solutions
of the characteristic equation \ $\det(\xi I-L)=0$. Solving these relations
with respect to $q$ we get the transformation $\lambda\rightarrow q$
\begin{equation}
q^{i}=\alpha_{i}(\lambda),\;\;\;\;\;i=1,...,n.\label{eqs}%
\end{equation}
The remaining part of the transformation to the separation coordinates can be
obtained as a canonical transformation reconstructed from the generating
function $W(p,\lambda)=\sum_{i}p_{i}\alpha_{i}(\lambda)$ in the standard way
by solving the implicit relations $\mu_{i}=\frac{\partial W(p,\lambda
)}{\partial\lambda^{i}}$ with respect to $p_{i}$ obtaining $p_{i}=\beta
_{i}(\lambda,\mu)$. In the $(\lambda,\mu)$ coordinates, known as the
Darboux-Nijenhuis coordinates (DN), the tensor $L$ is diagonal
$L=\mathrm{diag}(\lambda^{1},...,\lambda^{n})\equiv\Lambda_{n}$, while the
Hamiltonians (\ref{3.4}) of our quasi-bi-Hamiltonian chain attain the form
\cite{2}
\[
E_{r}(\lambda,\mu)=-\sum_{i=1}^{n}\frac{\partial\rho_{r}}{\partial\lambda^{i}%
}\frac{f_{i}(\lambda^{i})\mu_{i}^{2}}{\Delta_{i}}\;\;\;\;r=1,...,n,
\]
where
\[
\Delta_{i}=\prod\limits_{k=1,...n,\,k\neq i}(\lambda^{i}-\lambda^{k}),
\]
$\rho_{r}(\lambda)$ are symmetric polynomials (Vi\'{e}te polynomials) defined
by the relation
\begin{equation}
\det(\xi I-\Lambda)=(\xi-\lambda^{1})(\xi-\lambda^{2})...(\xi-\lambda
^{n})=\sum_{r=0}^{n}\rho_{r}\xi^{r},\label{3.10}%
\end{equation}
and where $f_{i}$ are arbitrary smooth functions of one real argument.

It turns out that there exists a sequence of generic separable potentials
$V_{r}^{(k)}$,\ $k\in\mathbf{Z},$ which can be added to geodesic Hamiltonians
$E_{r}$ such that the new Hamiltonians
\begin{equation}
H_{r}(q,p)=E_{r}(q,p)+V_{r}^{(k)}(q),\;\;\;\;r=1,...,n,\label{3,23}%
\end{equation}
are still separable in the same coordinates $(\lambda,\mu)$. These generic
potentials are given by some recursion relations \cite{5},\cite{8}. The
Hamiltonians $H_{r}:\mathcal{N}\rightarrow\mathbb{R}$ in (\ref{3,23}) satisfy
the following quasi-bi-Hamiltonian chain
\begin{equation}
\theta_{0}dH_{r+1}=\theta_{1}dH_{r}+\rho_{r}\theta_{0}dH_{1},\;\;\;\;H_{n+1}%
=0,\;\;\;\;\;\;r=1,...,n.\label{qBHV}%
\end{equation}
In the DN coordinates the Hamiltonians $H_{r}$ attain the form \cite{2}
\begin{equation}
H_{r}(\lambda,\mu)=-\sum_{i=1}^{n}\frac{\partial\rho_{r}}{\partial\lambda^{i}%
}\frac{f_{i}(\lambda^{i})\mu_{i}^{2}+\gamma_{i}(\lambda^{i})}{\Delta_{i}%
}\text{ },\;\;\;\;r=1,...,n,\label{pelnehamwlm}%
\end{equation}
where potentials $V_{r}^{(k)}$ enter $H_{r}$ as $\gamma_{i}(\lambda
^{i})=(\lambda^{i})^{n+k-1}$. From (\ref{pelnehamwlm}) it immediately follows
that in $(\lambda,\mu)$ variables the contravariant metric tensor $G$ and all
the Killing tensors $K_{r}$ are diagonal
\[
G^{ij}=\frac{f_{i}(\lambda^{i})}{\Delta_{i}}\delta^{ij},\;\;\;(K_{r})_{j}%
^{i}=-\frac{\partial\rho_{r}}{\partial\lambda^{i}}\delta_{j}^{i}.
\]
Moreover, in the $(\lambda,\mu)$ coordinates the recursion operator and the
tensor $\theta_{1}$ attain the form
\[
N=\left(
\begin{array}
[c]{cc}%
\Lambda_{n} & 0_{n}\\
0_{n} & \Lambda_{n}%
\end{array}
\right)  ,\;\;\;\theta_{1}=\left(
\begin{array}
[c]{cc}%
0_{n} & \Lambda_{n}\\
-\Lambda_{n} & 0_{n}%
\end{array}
\right)
\]
while $\theta_{0}$ remains in the form (\ref{3.1}) since the transformation
$(q,p)\rightarrow(\lambda,\mu)$ is canonical.

The quasi-bi-Hamiltonian chain (\ref{qBHV}) on $\mathcal{N}$ can easily be
lifted to a bi-Hamitonian chain (\ref{chain}) on the extended manifold
$\mathcal{M}=T^{\ast}Q\times\mathbf{R}=\mathcal{N}\times\mathbf{R}$
\cite{ibort}.

Having such a complete picture of separable quasi-bi-Hamiltonian chains on
Riemannian manifolds, one can ask a question: what kind of holonomic
constraints can be imposed on considered systems so that this
quasi-bi-Hamiltonian separability schema is preserved? The simplest admissible
case of such constraints will be considered in the next sections.

\section{Reduction $\lambda^{n}=0$}

Let us consider a particle moving in our Riemannian manifold $Q$ equipped with
the coordinates $(q^{1},...,q^{n})$. Suppose that this particle is
subordinated to some holonomic constraints on $Q$ defined by the set of
relations
\begin{equation}
\varphi_{k}(q)=0,\;\;k=1,...,r\label{redQ}%
\end{equation}
that define some submanifold of $Q.$ The velocity $v=\sum_{i}v^{i}%
\frac{\partial}{\partial q^{i}}$ of this particle must then remain tangent to
this submanifold so that%

\[
0=\left\langle d\varphi_{k},v\right\rangle =\sum_{i=1}^{n}\frac{\partial
\varphi_{k}}{\partial q^{i}}v^{i}.
\]
and thus in our coordinates $v^{i}=\sum_{j}G^{ij}p_{j}$ the motion of the
particle in the phase space $\mathcal{N}=T^{\ast}Q$ is constrained not only by
the $r$ relations (\ref{redQ}) but also by the $r$ relations%

\begin{equation}
\varphi_{r+k}(q,p)\equiv\sum_{i,j=1}^{n}G^{ij}(q)\frac{\partial\varphi_{k}%
(q)}{\partial q^{i}}p_{j}=0,\;\;\;\;\;\;\;k=1,...,r.\label{redM}%
\end{equation}
that are nothing else than the lift of (\ref{redQ}) to $\mathcal{N}$. We will
call the constraints (\ref{redM}) a $g$-\emph{consequence} of the constraints
(\ref{redQ}), as they are natural differential consequences of (\ref{redQ}) at
given metric tensor $g$. The constraints (\ref{redQ})-(\ref{redM}) define a
submanifold $\mathcal{S}$ of $\mathcal{N}$ of dimension $n-2r$.

Let us now consider our quasi-bi-Hamiltonian chain (\ref{qBHV}) in
$\mathcal{N}$. We would like to know what types of holonomic constraints
(\ref{redQ}) on $Q$ do not destroy the separability of the constrained chain?
This is a complicated question and due to its nature it is most convenient
(for a moment) to consider it directly in our separation coordinates
$(\lambda,\mu)$. Thus, in this section we will analyze only a very special
choice of the functions $\varphi_{k}$ in (\ref{redQ}). Namely, we put $r=1$
(so that $n-2r=n-2$ which corresponds to $m=2$ in Section 2) and define the
corresponding function $\varphi_{1}^{\prime}$ in $(\lambda,\mu)$ variables as
\begin{equation}
\varphi_{1}^{\prime}(\lambda)=\lambda^{n}\label{wiez2}%
\end{equation}
Since the metric tensor $G=(G^{ij})$ has in $(\lambda,\mu)$ coordinates the
diagonal form $G=\mathrm{diag}\left(  f_{i}(\lambda^{i})/\Delta_{i}\right)
\,,\,i=1,\ldots,n$ and since the equation (\ref{redM}) has an invariant form
(i.e. in $(\lambda,\mu)$ coordinates it has the same form with $q$ and $p$
replaced by $\lambda$ and $\mu$ respectively) the $g$-consequence of
(\ref{wiez2}) reads as
\begin{equation}
\varphi_{2}^{\prime}(\lambda,\mu)=\frac{f_{n}(\lambda^{n})}{\Delta_{n}}\mu
_{n}\label{wiez2a}%
\end{equation}
(we use $^{\prime}$ here since soon we will modify the constraints
(\ref{wiez2})-(\ref{wiez2a}) to a simpler form). These two constraints define
a subset $\mathcal{S}^{\prime}=\left\{  \varphi_{1}^{\prime}=0,\varphi
_{2}^{\prime}=0\right\}  $ of $\mathcal{N}$. We will however restrict us to a
submanifold $\mathcal{S}=\left\{  \lambda^{n}=0,\mu_{n}=0\right\}  $
neglecting the term $\ f_{n}(\lambda^{n})/\Delta_{n}$ in $\varphi_{2}^{\prime
}$ (for $\Delta_{n}\neq0$ $\mathcal{S}\subset\mathcal{S}^{\prime}$ and in case
of the systems when $f_{n}$ never vanishes $\mathcal{S}$ and $\mathcal{S}%
^{\prime}$ coincide). The reason for this is that various calculations with
the help of \ these new constraints $\lambda^{n}=0,\mu_{n}=0$ are simpler. Let
us now perform the Dirac reduction of our Poisson operators $\theta_{0}$ and
$\theta_{1}$%

\begin{equation}
\theta_{0}=\left(
\begin{array}
[c]{cc}%
0_{n} & I_{n}\\
-I_{n} & 0_{n}%
\end{array}
\right)  \text{ \ \ \ , \ \ }\theta_{1}=\left(
\begin{array}
[c]{cc}%
0_{n} & \Lambda_{n}\\
-\Lambda_{n} & 0_{n}%
\end{array}
\right) \label{formatet}%
\end{equation}
and of the corresponding quasi-bi-Hamiltonian chain (\ref{3.17}) (geodesic
case) or (\ref{qBHV}) (potential case) on the submanifold $\mathcal{S}$. We
can do it either by using the constraints (\ref{wiez2})-(\ref{wiez2a}) or by
the constraints that directly describe $\mathcal{S}$:
\begin{equation}
\text{ }\varphi_{1}(\lambda)\equiv\lambda^{n}=0\text{\ \ \ ,\ \ }\varphi
_{2}(\lambda,\mu)\equiv\mu_{n}=0\label{wiez1}%
\end{equation}
(of course $\varphi_{1}\equiv\varphi_{1}^{\prime}$). It turns out that the
corresponding Dirac deformations (given by (\ref{2.10}) or equivalently by
(\ref{2.17})) $\theta_{0,D}$ and $\theta_{0,D}^{\prime}$ of $\theta_{0}$ are
\emph{not} equal. Similarly, $\theta_{1,D}\neq\theta_{1,D}^{\prime}$. However,
after reducing $\theta_{0,D}$ and $\theta_{0,D}^{\prime}$ on $\mathcal{S}$ we
obtain the same Poisson operator (and similarly with $\theta_{1,D}$ and
$\theta_{1,D}^{\prime}$).

\begin{theorem}
With the notation as above, the reduced operators $\theta_{i,R}$ and
$\theta_{i,R}^{\prime}$ on $\mathcal{S}$ satisfy $\theta_{0,R}^{\prime}%
=\theta_{0,R}$ and $\theta_{1,R}^{\prime}=\theta_{1,R}$. If we parametrize
$\mathcal{S}$ by the variables $(\lambda^{1},\ldots,\lambda^{n-1},\mu
_{1},\ldots,\mu_{n-1})$ then
\begin{equation}
\theta_{0,R}=\left(
\begin{array}
[c]{cc}%
0_{n-1} & I_{n-1}\\
-I_{n-1} & 0_{n-1}%
\end{array}
\right)  \text{ \ \ , \ }\theta_{1,R}=\left[
\begin{array}
[c]{cc}%
0_{n-1} & \Lambda_{n-1}\\
-\Lambda_{n-1} & 0_{n-1}%
\end{array}
\right]  \text{\ \ }\label{zredukowanetety}%
\end{equation}
with $\Lambda_{n-1}=\mathrm{diag}(\lambda^{1},\ldots,\lambda^{n-1})$.
\end{theorem}

\begin{proof}
Let us start by calculating $\theta_{0,D}$. In this case the $2\times2$ Gram
matrix $\varphi=(\left\{  \varphi_{i},\varphi_{j}\right\}  _{\theta_{0}})$ is
of the form
\[
\varphi=\left[
\begin{array}
[c]{cc}%
0 & 1\\
-1 & 0
\end{array}
\right]
\]
and thus according to (\ref{2.8}) we have $Z_{1}=X_{2}$ and $Z_{2}=-X_{1}$.
The formula (\ref{2.10}) yields
\[
\theta_{0,D}=\theta_{0}-X_{1}\wedge X_{2}=\theta_{0}-\frac{\partial}%
{\partial\lambda^{n}}\wedge\frac{\partial}{\partial\mu_{n}}=\sum
\limits_{i=1}^{n-1}\frac{\partial}{\partial\lambda^{i}}\wedge\frac{\partial
}{\partial\mu_{i}},
\]

since $X_{1}=\theta_{0}d\varphi_{1}=-\partial/\partial\mu_{n}$ and
$X_{2}=\theta_{0}d\varphi_{2}=\partial/\partial\lambda^{n}$. So, in
$(\lambda,\mu)$ coordinates we have
\begin{equation}
\theta_{0,D}=\left[
\begin{array}
[c]{c|c}%
0_{n} &
\begin{array}
[c]{cc}%
I_{n-1} & 0_{(n-1)\times1}\\
0_{1\times(n-1)} & 0
\end{array}
\\\hline%
\begin{array}
[c]{cc}%
-I_{n-1} & 0_{(n-1)\times1}\\
0_{1\times(n-1)} & 0
\end{array}
& 0_{n}%
\end{array}
\right] \label{teta0D}%
\end{equation}

The operator (\ref{teta0D}) has thus the $n$-th and the last column and the
$n$-th and the last row filled with zeros. These two rows and two columns of
zeros correspond to the fact (see Theorem \ref{obciecie}) that the constraints
$\varphi_{1}=\lambda^{n}$ and $\varphi_{2}=\mu_{n}$ are now Casimir functions
for $\theta_{0,D}$. We can thus directly project $\theta_{0,D}$ onto the
symplectic leaf $\mathcal{S}=\left\{  \lambda^{n}=\mu_{n}=0\right\}  $ of
$\theta_{0,D}$ by simply removing the two zero columns and zero rows from
(\ref{teta0D}), which yield the operator $\theta_{0,R}$ that written in
coordinates $(\lambda^{1},\ldots\lambda^{n-1},\mu_{1},\ldots\mu_{n-1})$ on
$\mathcal{S}$ attains the form as in (\ref{zredukowanetety}). Similar
computations show that in the case of $\theta_{1,D}$%
\[
\varphi=\left[
\begin{array}
[c]{cc}%
0 & \lambda^{n}\\
-\lambda^{n} & 0
\end{array}
\right]  =\left[
\begin{array}
[c]{cc}%
0 & \varphi_{1}\\
-\varphi_{1} & 0
\end{array}
\right]
\]
and thus (see (\ref{first})) $\varphi_{1}$ and $\varphi_{2}$ are first-class
constraints. Indeed, following (\ref{2.8}) we have $Z_{1}=\frac{1}{\lambda
^{n}}X_{2}$ and $Z_{2}=-\frac{1}{\lambda^{n}}X_{1}$ and the formula
(\ref{2.10}) yields
\[
\theta_{1,D}=\theta_{1}-\frac{1}{\lambda^{n}}X_{1}\wedge X_{2}%
\]
which seems to be singular on $\mathcal{S}.$ Nevertheless this singularity is
'removable' since $X_{1}=\theta_{1}d\varphi_{1}=-\lambda^{n}\partial
/\partial\mu_{n}$, $X_{2}=\theta_{1}d\varphi_{2}=\lambda^{n}\partial
/\partial\lambda^{n}$ and hence
\[
\theta_{1,D}=\theta_{1}-\lambda^{n}\frac{\partial}{\partial\lambda^{n}}%
\wedge\frac{\partial}{\partial\mu_{n}}=\sum\limits_{i=1}^{n-1}\lambda^{i}%
\frac{\partial}{\partial\lambda^{i}}\wedge\frac{\partial}{\partial\mu_{i}}.
\]
So, the matrix form of $\theta_{1,D}$ in $(\lambda,\mu)$ coordinates becomes
\[
\theta_{1,D}=\left[
\begin{array}
[c]{c|c}%
0_{n} &
\begin{array}
[c]{cc}%
\Lambda_{n-1} & 0_{(n-1)\times1}\\
0_{1\times(n-1)} & 0
\end{array}
\\\hline%
\begin{array}
[c]{cc}%
-\Lambda_{n-1} & 0_{(n-1)\times1}\\
0_{1\times(n-1)} & 0
\end{array}
& 0_{n}%
\end{array}
\right]
\]
and the projection of this operator onto the symplectic leaf $\mathcal{S}%
=\left\{  \lambda^{n}=\mu_{n}=0\right\}  $ of $\theta_{1,D}$ yields exactly
$\theta_{1,R}$ as in (\ref{zredukowanetety}).

Passing to $\varphi_{1}^{\prime},\varphi_{2}^{\prime}$ notice that
\[
\varphi_{12}^{\prime}=\{\varphi_{1}^{\prime},\varphi_{2}^{\prime}%
\}_{\theta_{0}}=\frac{f_{n}(\lambda^{n})}{\Delta_{n}},
\]
so one can have a situation when $\varphi_{12}^{\prime}|_{\mathcal{S}}=0$
which again leads to first-class constraints according to Dirac
classification. Nevertheless, calculations similar to these for $\varphi
_{1},\varphi_{2}$ show that singularities are again 'removable' and we end up
with the same form of $\theta_{0,D}$ and $\theta_{1,D}$.
\end{proof}

Of course, the reduced operators $\theta_{0,R}$ and $\theta_{1,R}$ are
compatible. Moreover, it is easy to see that the Hamiltonians
(\ref{pelnehamwlm}) restricted to $\mathcal{S}$ become
\begin{align*}
&  H_{r,R}(\lambda^{1},\ldots\lambda^{n-1},\mu_{1},\ldots\mu_{n-1})\\
&  =H_{r}|_{\mathcal{S}}=-\sum_{i=1}^{n-1}\frac{\partial\rho_{r,R}}%
{\partial\lambda^{i}}\frac{f_{i,R}(\lambda^{i})\mu_{i}^{2}+\gamma
_{i,R}(\lambda^{i})}{\Delta_{i,R}},\text{ \ \ }r=1,...,n-1,
\end{align*}%
\[
H_{n,R}=H_{n}|_{\mathcal{S}}=0
\]
where $f_{i,R}(\lambda^{i})=f_{i}(\lambda^{i})/\lambda^{i}$ , $\gamma
_{i,R}(\lambda^{i})=\gamma_{i}(\lambda^{i})/\lambda^{i},$ $i=1,\ldots,n-1$ and
$\rho_{r,R}(\lambda)$ are Vi\'{e}te polynomials of dimension $n-1 $. Having
all this is mind, we can formulate the following corollary.

\begin{corollary}
Both Dirac reductions (\ref{wiez2})-(\ref{wiez2a}) and (\ref{wiez1}) of the
quasi-bi-Hamiltonian system (\ref{qBHV}) onto $\mathcal{S}$ lead to the same
quasi-bi-Hamiltonian system of the form
\[
\theta_{0,R}dH_{r+1,R}=\theta_{1,R}dH_{r,R}+\rho_{r,R}\theta_{0,R}%
dH_{1,R},\;\;\;\;H_{n,R}=0,\;\;\;\;\;\;r=1,...,n-1.
\]
which is thus separable in the variables $(\lambda^{1},\ldots\lambda^{n-1}%
,\mu_{1},\ldots\mu_{n-1})$ that coincide on $\mathcal{S}$ with 'first' $n-2$
separation coordinates from our coordinate system $(\lambda,\mu)$ that
separates (\ref{qBHV}).
\end{corollary}

Because the reduced quasi-bi-Hamiltonian chain has exactly the form
(\ref{qBHV}), it can be put into a bi-Hamiltonian chain on $\mathcal{S\times
}\mathbb{R.}$

Of course this procedure of Dirac reduction, in principle could be performed
directly in $(q,p)$ coordinates, since the formula (\ref{2.10}) has a tensor
character and thus yields the same result no matter what coordinate system we
choose for actual calculating of deformations $\theta_{0,D}$ and $\theta
_{1,D}$. There is however a major obstacle here: we are usually not able to
express the constraint $\lambda^{n}=0$ in 'physical' coordinates $(q,p)$, as
equations (\ref{eqs}) are noninvertible in general, i.e. there is no algebraic
way of solving them with respect to $\lambda$ (if we could, then the second
constraint (\ref{wiez2a}) could be computed as the $g$-consequence of the
first one calculated directly in $(q,p)$ coordinates, with the help of
(\ref{redM})). Thus, although the picture presented in this section is clear,
nevertheless it is somehow useless once we are given the quasi-bi-Hamiltonian
chain in natural $(q,p)$ coordinates. In the next section we will demonstrate
how this problem can be defused by the use of the so called Hankel-Frobenius coordinates.

\section{Reduction in Hankel-Frobenius coordinates}

As it was demonstrated in the previous section, the constraints $\varphi
_{1}=\lambda^{n},$ $\varphi_{2}=\mu_{n}$ preserve the separability on the
constrained submanifold $\mathcal{S}$ but are very inconvenient to handle
with, as in general we do not know how to express them in the original
coordinates $(q,p).$ More convenient for this purpose is the set of so called
Hankel-Frobenius coordinates $(\rho_{i},v_{i})_{i=1}^{n}$ \cite{HF1}%
,\cite{HF2}. They are non-canonical coordinates, related to the separated
coordinates in the following way
\begin{equation}%
\begin{array}
[c]{l}%
\rho_{i}=\rho_{i}(\lambda),\ \ \ \ \ \ \ i=1,...,n,\\
v_{i}=%
%TCIMACRO{\dsum \limits_{j}}%
%BeginExpansion
{\displaystyle\sum\limits_{j}}
%EndExpansion
(V_{n}^{-1})_{ij}\mu_{j}=%
%TCIMACRO{\dsum \limits_{k}}%
%BeginExpansion
{\displaystyle\sum\limits_{k}}
%EndExpansion
\dfrac{\partial\rho_{i}}{\partial\lambda^{k}}\dfrac{1}{\Delta_{k}}\mu
_{k},\ \ \ \ \ i=1,...,n,
\end{array}
\label{HF}%
\end{equation}
where
\[
V_{n}=\left(
\begin{array}
[c]{cccc}%
(\lambda^{1})^{n-1} & \cdots & \lambda^{1} & 1\\
\vdots & \vdots & \vdots & \vdots\\
(\lambda^{n})^{n-1} & \cdots & \lambda^{n} & 1
\end{array}
\right)
\]
is a Vandermonde matrix and $\rho_{i}$ are Vi\'{e}te polynomials (\ref{3.10}).
The mapping (\ref{HF}) is not a point transformation on $Q$ and therefore it
makes no sense to distinguish covariant and contravariant indices now.
Applying the map (\ref{HF}) we obtain the form of operators $\theta_{0}$ and
$\theta_{1}$ and the $2n$-dimensional recursion operator $N_{n}$ (\ref{3.14})
in $(\rho,v)$ coordinates: \
\begin{equation}
\theta_{0}=\left(
\begin{array}
[c]{cc}%
0 & U_{n}\\
-U_{n} & 0
\end{array}
\right)  ,\ \ \ \ \ \ \ U_{n}=\left(
\begin{array}
[c]{cccc}%
0 & 0 & \cdots & 1\\
0 & \cdots & 1 & \rho_{1}\\
\vdots & \cdots & \cdots & \vdots\\
1 & \rho_{1} & \cdots & \rho_{n-1}%
\end{array}
\right)  ,\label{5.2}%
\end{equation}%
\begin{equation}
\theta_{1}=N_{n}\theta_{0}=\left(
\begin{array}
[c]{cc}%
0 & F_{n}U_{n}\\
-F_{n}U_{n} & 0
\end{array}
\right)  ,\ \ \ \ \ \ \ F_{n}U_{n}=U_{n}F_{n}^{T}.\label{5.4}%
\end{equation}%
\[
N_{n}=\left(
\begin{array}
[c]{cc}%
F_{n} & 0\\
0 & F_{n}%
\end{array}
\right)  ,\ \ \ \ \ \ \ \ F_{n}=\left(
\begin{array}
[c]{ccccc}%
-\rho_{1} & 1 & \cdots & \cdots & 0\\
-\rho_{2} & 0 & 1 & \cdots & 0\\
\vdots & \vdots &  &  & \vdots\\
-\rho_{n-1} & 0 & \cdots & \cdots & 1\\
-\rho_{n} & 0 & \cdots & \cdots & 0
\end{array}
\right)  .
\]
We will now consider yet another deformation (\ref{2.10}) of $\theta_{0}$ and
$\theta_{1}$, given in $(\lambda,\mu)$ variables by the constraints
\begin{align}
\varphi_{1}^{\prime\prime} &  =\rho_{n}(\lambda)=(-1)^{n}\lambda^{1}%
\lambda^{2}\cdots\lambda^{n}\label{wiez3}\\
\varphi_{2}^{\prime\prime} &  =v_{n}(\lambda)=(-1)^{n}%
%TCIMACRO{\dsum \limits_{j=1}^{n}}%
%BeginExpansion
{\displaystyle\sum\limits_{j=1}^{n}}
%EndExpansion
\frac{\mu_{j}}{\Delta_{j}}\lambda^{1}\lambda^{2}\cdots\lambda^{j-1}%
\lambda^{j+1}\cdots\lambda^{n}\nonumber
\end{align}
equal to last pair of Hankel-Frobenius coordinates. The constraints
(\ref{wiez3}) are related with the constraints (\ref{wiez1}) as
\begin{align}
\varphi_{1}^{\prime\prime} &  =\psi_{1}\varphi_{1}\label{deform13}\\
\varphi_{2}^{\prime\prime} &  =\psi_{2}\varphi_{1}+\psi_{3}\varphi
_{2}\nonumber
\end{align}
where the functions
\begin{align*}
\psi_{1} &  =(-1)^{n}\lambda^{1}\lambda^{2}\cdots\lambda^{n-1}\text{ \ }\\
\text{\ \ \ }\psi_{2} &  =(-1)^{n}%
%TCIMACRO{\dsum \limits_{j=1}^{n-1}}%
%BeginExpansion
{\displaystyle\sum\limits_{j=1}^{n-1}}
%EndExpansion
\frac{\mu_{j}}{\Delta_{j}}\lambda^{1}\lambda^{2}\cdots\lambda^{j-1}%
\lambda^{j+1}\cdots\lambda^{n-1}%
\end{align*}
and $\psi_{3}=(-1)^{n}\lambda^{1}\lambda^{2}\cdots\lambda^{n-1}/\Delta_{n}$
never vanish at $\mathcal{S}=\left\{  \varphi_{1}=\varphi_{2}=0\right\}  $
\ (in fact, $\psi_{2}|_{\mathcal{S}}=-1$ and $\psi_{3}|_{\mathcal{S}}%
=-(\mu_{1}+\ldots+\mu_{n-1})$ ) and thus the constraints (\ref{wiez3}) define
(locally) the same submanifold $\mathcal{S}$ as the constraints (\ref{wiez1}).
The corresponding deformations $\theta_{0,D}^{\prime\prime}$ and $\theta
_{1,D}^{\prime\prime}$ of $\theta_{0} $ and $\theta_{1}$ will of course not be
equal to $\theta_{0,D}$ and $\theta_{1,D}$, but again it turns out that their
reductions on $S$ will coincide with the corresponding reductions of
$\theta_{0,D}$ and $\theta_{1,D}$ on $\mathcal{S}$.

\begin{theorem}
\label{red1=red3}In the notation as above, $\theta_{0,R}^{\prime\prime}%
=\theta_{0,R}$ and $\theta_{0,R}^{\prime\prime}=\theta_{0,R}$
\end{theorem}

\begin{proof}
We will use (\ref{2.17}) rather than (\ref{2.10}) since it turns out that the
calculations are in this case simpler when one uses bracket definition of
Dirac deformation than bivector definition. Applying (\ref{2.17}), we easily
get that for any two functions $A,B:\mathcal{N}\rightarrow R$
\[
\left\{  A,B\right\}  _{\theta_{0,D}}=\left\{  A,B\right\}  _{\theta_{0}%
}+\frac{\left\{  A,\varphi_{2}\right\}  _{\theta_{0}}\left\{  B,\varphi
_{1}\right\}  _{\theta_{0}}-\left\{  A,\varphi_{1}\right\}  _{\theta_{0}%
}\left\{  B,\varphi_{2}\right\}  _{\theta_{0}}}{\left\{  \varphi_{1}%
,\varphi_{2}\right\}  _{\theta_{0}}}%
\]
where of course $\left\{  \varphi_{1},\varphi_{2}\right\}  _{\theta_{0}}=1$
and so it does not vanish on $\mathcal{S}$. Similarly
\begin{equation}
\left\{  A,B\right\}  _{\theta_{0,D}^{\prime\prime}}=\left\{  A,B\right\}
_{\theta_{0}}+\frac{\left\{  A,\varphi_{2}^{\prime\prime}\right\}
_{\theta_{0}}\left\{  B,\varphi_{1}^{\prime\prime}\right\}  _{\theta_{0}%
}-\left\{  A,\varphi_{1}^{\prime\prime}\right\}  _{\theta_{0}}\left\{
B,\varphi_{2}^{\prime\prime}\right\}  _{\theta_{0}}}{\left\{  \varphi
_{1}^{\prime\prime},\varphi_{2}^{\prime\prime}\right\}  _{\theta_{0}}%
}.\label{modyfbis}%
\end{equation}
Using the relations (\ref{deform13}) between the deformed constraints
$\varphi_{i}^{\prime\prime}$ and the original constraints $\varphi_{i}$, the
Leibniz property of Poisson brackets and the fact that $\psi_{1}$ and
$\psi_{3}$ depend only on $\lambda$, we obtain
\begin{align*}
\left\{  \varphi_{1}^{\prime\prime},\varphi_{2}^{\prime\prime}\right\}
_{\theta_{0}} &  =\psi_{1}\varphi_{1}\left\{  \varphi_{1},\psi_{2}\right\}
_{\theta_{0}}+\varphi_{1}^{2}\left\{  \psi_{1},\psi_{2}\right\}  _{\theta_{0}%
}+\\
&  +\psi_{1}\psi_{3}\left\{  \varphi_{1},\varphi_{2}\right\}  _{\theta_{0}%
}+\varphi_{1}\varphi_{2}\left\{  \psi_{1},\psi_{3}\right\}  _{\theta_{0},}%
\end{align*}
so that $\left\{  \varphi_{1}^{\prime\prime},\varphi_{2}^{\prime\prime
}\right\}  _{\theta_{0}}|_{\mathcal{S}}=\psi_{1}\psi_{3}\left\{  \varphi
_{1},\varphi_{2}\right\}  _{\theta_{0}}|_{\mathcal{S}}=\psi_{1}\psi
_{3}|_{\mathcal{S}}$. Similar calculations show that
\begin{align*}
&  \left.  \left(  \left\{  A,\varphi_{2}^{\prime\prime}\right\}  _{\theta
_{0}}\left\{  B,\varphi_{1}^{\prime\prime}\right\}  _{\theta_{0}}-\left\{
A,\varphi_{1}^{\prime\prime}\right\}  _{\theta_{0}}\left\{  B,\varphi
_{2}^{\prime\prime}\right\}  _{\theta_{0}}\right)  \right\vert _{\mathcal{S}%
}\\
&  =\psi_{1}\psi_{3}\left.  \left(  \left\{  A,\varphi_{2}\right\}
_{\theta_{0}}\left\{  B,\varphi_{1}\right\}  _{\theta_{0}}-\left\{
A,\varphi_{1}\right\}  _{\theta_{0}}\left\{  B,\varphi_{2}\right\}
_{\theta_{0}}\right)  \right\vert _{\mathcal{S}}%
\end{align*}
and thus the factors $\psi_{1}\psi_{3}$ in the numerator and in the
denominator of (\ref{modyfbis}) cancel and we conclude that
\[
\left.  \left\{  A,B\right\}  _{\theta_{0,D}^{\prime\prime}}\right\vert
_{\mathcal{S}}=\left.  \left\{  A,B\right\}  _{\theta_{0,D}}\right\vert _{S}%
\]
which is the same as to claim that $\theta_{0,R}^{\prime\prime}=\theta_{0,R}$.
The proof that $\theta_{1,R}^{\prime\prime}=\theta_{1,R}$ is similar: first
one shows that
\[
\left\{  \varphi_{1}^{\prime\prime},\varphi_{2}^{\prime\prime}\right\}
_{\theta_{1}}|_{\mathcal{S}}=\psi_{1}\psi_{3}\left\{  \varphi_{1},\varphi
_{2}\right\}  _{\theta_{1}}|_{\mathcal{S}}=\psi_{1}\psi_{3}|_{\mathcal{S}%
}\,\lambda^{n}%
\]
which is by the way equal to zero on $\mathcal{S}$. However , we also get
\begin{align*}
&  \left.  \left(  \left\{  A,\varphi_{2}^{\prime\prime}\right\}  _{\theta
_{1}}\left\{  B,\varphi_{1}^{\prime\prime}\right\}  _{\theta_{1}}-\left\{
A,\varphi_{1}^{\prime\prime}\right\}  _{\theta_{1}}\left\{  B,\varphi
_{2}^{\prime\prime}\right\}  _{\theta_{1}}\right)  \right\vert _{\mathcal{S}%
}\\
&  =\left.  \psi_{1}\psi_{3}\left(  \left\{  A,\varphi_{2}\right\}
_{\theta_{1}}\left\{  B,\varphi_{1}\right\}  _{\theta_{1}}-\left\{
A,\varphi_{1}\right\}  _{\theta_{1}}\left\{  B,\varphi_{2}\right\}
_{\theta_{1}}\right)  \right\vert _{\mathcal{S}}%
\end{align*}
so that on $\mathcal{S}$
\[
\left.  \left\{  A,B\right\}  _{\theta_{1,D}}^{\prime\prime}\right\vert
_{\mathcal{S}}=\left.  \left\{  A,B\right\}  _{\theta_{1}}\right\vert
_{\mathcal{S}}+\left.  \frac{\psi_{1}\psi_{3}\left(  \left\{  A,\varphi
_{2}\right\}  _{\theta_{1}}\left\{  B,\varphi_{1}\right\}  _{\theta_{1}%
}-\left\{  A,\varphi_{1}\right\}  _{\theta_{1}}\left\{  B,\varphi_{2}\right\}
_{\theta_{1}}\right)  }{\psi_{1}\psi_{3}\left\{  \varphi_{1},\varphi
_{2}\right\}  _{\theta_{1}}}\right\vert _{\mathcal{S}}%
\]%
\[
=\left.  \left\{  A,B\right\}  _{\theta_{1}}\right\vert _{\mathcal{S}}+\left.
\frac{\left\{  A,\varphi_{2}\right\}  _{\theta_{1}}\left\{  B,\varphi
_{1}\right\}  _{\theta_{1}}-\left\{  A,\varphi_{1}\right\}  _{\theta_{1}%
}\left\{  B,\varphi_{2}\right\}  _{\theta_{1}}}{\lambda^{n}}\right\vert
_{\mathcal{S}}=\left.  \left\{  A,B\right\}  _{\theta_{1,D}}\right\vert
_{\mathcal{S}}%
\]
since the term $\lambda^{n}$ in the last expression does not cause any
singularity: for every possible combination $A,B=\lambda^{i},\mu_{j}$ the
numerator in the above expression is either zero or some multiple of
$\lambda^{n}$. This proves that $\theta_{1,R}^{\prime\prime}=\theta_{1,R}.$
\end{proof}

Before we proceed with the main theme of this article, let us make a
digression: we will establish the form of $\theta_{0,R}$ and $\theta_{1,R}$ in
$(\rho,v)$-coordinates.

\begin{lemma}
In $(\rho,v)$-coordinates we have
\begin{align*}
\theta_{0,R}  &  =\left(  N_{n-1}\right)  ^{-1}\left(
\begin{array}
[c]{cc}%
0 & U_{n-1}\\
-U_{n-1} & 0
\end{array}
\right)  =\left(
\begin{array}
[c]{cc}%
0_{n-1} & \left(  F_{n-1}\right)  ^{-1}U_{n-1}\\
-\left(  F_{n-1}\right)  ^{-1}U_{n-1} & 0_{n-1}%
\end{array}
\right) \\
\theta_{1,R}  &  =\left(
\begin{array}
[c]{cc}%
0 & U_{n-1}\\
-U_{n-1} & 0
\end{array}
\right)
\end{align*}

\end{lemma}

This lemma can be proved either by direct calculating of the deformation
formula (\ref{2.10}) in $(\rho,v)$-coordinates or by transforming both
$\theta_{0,R}$ and $\theta_{1,R}$ with the map that is the restriction of the
map (\ref{HF}) to $\mathcal{S}$. This restricted map has no longer the form
(\ref{HF}) and this is why the operators $\theta_{0,R}$ and $\theta_{1,R}$
does not transform respectively to the operators of the form (\ref{5.2}) and
(\ref{5.4}) with $n$ replaced by $n-1$.

Having established Theorem \ref{red1=red3} we can now, for a given
(quasi)-bi-Hamiltonian system (\ref{qBHV}), perform the Dirac reduction
$\lambda^{n}=\mu_{n}=0$ directly in physical coordinates $(q,p)$ by performing
the equivalent reduction $\rho_{n}=v_{n}=0$ provided that we are able to
express the constraints (\ref{wiez3}) directly in physical coordinates. The
first of the constraints in (\ref{wiez3}), i.e. $\rho_{n}=0 $ is given
directly in $(q,p)$ coordinates once the system (\ref{3.17}) has been given
(this is the reason for which we consider this constraint and Hankel-Frobenius
coordinates). Below we shall show how to express the second constraint
($v_{n}=0$) in physical coordinates.

Let us first observe that in case of systems for which $f_{i}(\lambda
^{i})=const$ (that does not depend on $i$) the constraint $v_{n}=0$ calculated
in $(\lambda,\mu)$ coordinates is just the $g$-consequence of the first one
($\rho_{n}=0$) , so that in this particular case we can easily get the
function $\phi_{2}^{\prime\prime}(q,p)$ by calculating the expression
(\ref{redM}) with an appropriate metric tensor $g$ that can be found in
$h_{1}$. When the functions $f_{i}$ are more complicated, we must proceed differently.

\begin{theorem}
Assume that $f_{i}(\lambda^{i})=f(\lambda^{i})$, i.e. that the functions
$f_{i}$ do not depend on $i$. Let also $F$ be the matrix given by (\ref{5.4}).
Then the coordinates $v_{i}$ expressed in $(q,p)$ coordinates have the
following form
\begin{equation}
v_{i}(q,p)=\sum\limits_{j=1}^{n}\left(  {}\left(  f(F)^{-1}\right)  _{ij}%
\sum_{k,l=1}^{n}G^{kl}(q)\frac{\partial\rho_{j}}{\partial q^{k}}p_{l}\right)
\label{nuwqp}%
\end{equation}
with $F$ being expressed in $q.$
\end{theorem}

\begin{proof}
Let us expand the function $f$ in a formal Laurent series
\begin{equation}
f(\lambda)=\sum\limits_{m}a_{m}\lambda^{m}\text{\thinspace, \ }a_{m}%
\in\mathbb{R}\text{ \ , \ }m\in\mathbf{Z}_{f}\subset\mathbf{Z.}\label{Laurent}%
\end{equation}
Let also $v_{k}^{\prime}$ , $k=1,\ldots,n$ be the $g$-consequence of $\rho
_{k}$ so that
\begin{equation}
v_{k}^{\prime}(\lambda,\mu)=\sum\limits_{i=1}^{n}\frac{f_{i}}{\Delta_{i}}%
\frac{\partial\rho_{k}}{\partial\lambda^{i}}\mu_{i}\text{ and }v_{k}^{\prime
}(q,p)=\sum\limits_{i,j=1}^{n}G^{ij}(q)\frac{\partial\rho_{k}}{\partial q^{i}%
}p_{j}.\label{nuprim}%
\end{equation}
Setting $\mu=(\mu_{1},\ldots,\mu_{n})^{T}$, $v=(v_{1},\ldots,v_{n})^{T}$,
$v^{\prime}=(v_{1}^{\prime},\ldots,v_{n}^{\prime})^{T}$ and using the fact
(\ref{HF}) that $\mu=Vv$ (where $V$ is the Vandermonde matrix) we easily
obtain that
\[
v^{\prime}=V^{-1\,}\mathrm{diag}(f(\lambda^{1}),\ldots,f(\lambda^{n}))\,V\,v
\]

Moreover, from the transformation between $(\lambda,\mu)$ and $(\rho,v)$
coordinates one finds that
\[
V^{-1}\Lambda^{m}V=U(F^{T})^{m}U^{-1}=F^{m}.
\]
so that due to (\ref{Laurent})
\begin{align*}
v^{\prime}  &  =V^{-1}\sum\limits_{m}a_{m}\text{ }\mathrm{diag}\left(
(\lambda^{1})^{m},\ldots,(\lambda^{n})^{m}\right)  \text{ }V\,v=V^{-1}\left(
\sum\limits_{m}a_{m}\text{ }\Lambda^{m}\right)  V\,v\\
&  =\left(  \sum\limits_{m}a_{m}\text{ }V^{-1}\Lambda^{m}V\right)  \,v=\left(
\sum\limits_{m}a_{m}F^{m}\right)  \,v=f(F)v.
\end{align*}
This relation is valid in every coordinate system and thus
$v(q,p)=f(F(q))v^{\prime}(q,p)$ which due to (\ref{nuprim}) yields
(\ref{nuwqp}).
\end{proof}

In case when the assumption of the theorem is not satisfied, we can not
express $v_{n}$ in $(q,p)$ variables using the method presented in the proof.
Since in practice this situation is very rare, we choose not to discuss it in
this article. Notice that if $f=const$ then $f(F)=const\,I$ and the formula
(\ref{nuwqp}) for $v_{n}$ reduces to the $g$-consequence of $\rho_{n}$, as it
should be.

\begin{remark}
\label{dlaczego} One can ask the question why, in the case of $f\neq const$,
instead of taking $\varphi_{2}^{\prime\prime}=\nu_{n}$ as above, we do not
choose $\varphi_{2}^{\prime\prime}$ to be simply the $g$-consequence of
$\varphi_{1}^{\prime\prime}=\rho_{n}$, as both pairs of constraints describe
the same $\mathcal{S}$. The answer is due to one of the fundamental
observations of presented paper, mentioned at the end of section 2. Actually,
a constrained submanifold $\mathcal{S}$ can be defined by infinitely many
different pairs of constraints, all of them of the first class according to
the classical Dirac classification. That is, either the Gram matrix is
singular: $\left\{  \varphi_{1}^{\prime\prime},\varphi_{2}^{\prime\prime
}\right\}  _{\theta}=0,$ or $\left\{  \varphi_{1}^{\prime\prime},\varphi
_{2}^{\prime\prime}\right\}  _{\theta}|_{\mathcal{S}}=0,$ for one or both
$\theta_{i},$ hence, in principle, there is no Dirac deformation $\theta
_{i,D}$\ or the restriction of $\theta_{i,D} $\ on $\mathcal{S}$ is not
possible. However, among all these pairs of constraints, there are exceptional
pairs, like $(\lambda^{n},\mu_{n})$ or $(\rho_{n},\nu_{n}),$ for which
singularities are 'removable' and Dirac restricted Poisson pencil is
nonsingular. So in fact, these particular constraints have to be considered as
second order constraints.
\end{remark}

Notice also that in order to determine the constraint $v_{n}$ from
(\ref{nuwqp}) we have first to find the function $f$ for example from $E_{1},$
written in $(\lambda,\mu)$ coordinates. So in practice, we cannot avoid the
calculation of separation coordinates, and this is the price we have to pay.
Nevertheless, in general, it is not difficult to find the transformation
(\ref{eqs}) and then the geodesic Hamiltonian $E_{1}(\lambda,\mu)$ for the
original system. Also, one has to bear in mind that even though the reduction
procedure looks trivial in $(\lambda,\mu)$-coordinates, it is usually not
possible to express the obtained reduced system back in $(q,p)$-coordinates
since our Hamiltonians $E_{r,R}$ have different function $f$ than the
Hamiltonians $E_{r}$. That is why we had to use Hankel-Frobenius coordinates.

\section{Examples}

In this chapter we will illustrate the introduced ideas by two examples. In
our first example of Dirac reductions of separable systems we will consider
the so called first Newton representation of the seventh-order stationary flow
of the KdV hierarchy \cite{myPhysicaA},\cite{1}. It is a Lagrangian system of
second order Newton equations
\begin{align*}
q_{,tt}^{1} &  =-10(q^{1})^{2}+4q^{2}\\
q_{,tt}^{2} &  =-16q^{1}q^{2}+10(q^{1})^{3}+4q^{3}\\
q_{,tt}^{3} &  =-20q^{1}q^{3}-8(q^{2})^{2}+30(q^{1})^{2}q^{2}-15(q^{1})^{4}+c
\end{align*}
(where the subscript $,t$ denotes the differentiation with respect to the
evolution parameter $t\in\mathbb{R}$). The above system can be represented as
a quasi-bi-Hamiltonian system on $\mathcal{N}$ belonging to a
quasi-bi-Hamiltonian chain of the form (\ref{qBHV}) with $n=3$, with Hamiltonians%

\begin{align}
H_{1} &  =p_{1}p_{3}+\tfrac{1}{2}p_{2}^{2}+10(q^{1})^{2}q^{3}-4q^{2}%
q^{3}+8q^{1}(q^{2})^{2}-10(q^{1})^{3}q^{2}+3(q^{1})^{5}\\
H_{2} &  =\tfrac{1}{2}q^{3}p_{3}^{2}-\tfrac{1}{2}q^{1}p_{2}^{2}+\tfrac{1}%
{2}q^{2}p_{2}p_{3}-\tfrac{1}{2}p_{1}p_{2}-\tfrac{1}{2}q^{1}p_{1}p_{3}%
+2(q^{1})^{2}(q^{2})^{2}+\tfrac{5}{2}(q^{1})^{4}q^{2}\nonumber\\
&  -\tfrac{5}{4}(q^{1})^{6}-2(q^{2})^{3}+(q^{3})^{2}-6q^{1}q^{2}%
q^{3}\label{hamex1}\\
H_{3} &  =\tfrac{1}{8}(q^{2})^{2}p_{3}^{2}+\tfrac{1}{8}(q^{1})^{2}p_{2}%
^{2}+\tfrac{1}{8}p_{1}^{2}+\tfrac{1}{4}q^{1}p_{1}p_{2}+\tfrac{1}{4}q^{2}%
p_{1}p_{3}-\tfrac{1}{4}q^{1}q^{2}p_{2}p_{3}\nonumber\\
&  -\tfrac{1}{2}q^{3}p_{2}p_{3}-3(q^{1})^{3}(q^{2})^{2}+q^{1}(q^{2}%
)^{3}+\tfrac{5}{4}(q^{1})^{5}q^{2}+2q^{1}(q^{3})^{2}\nonumber\\
&  +\tfrac{5}{4}(q^{1})^{4}q^{3}+(q^{2})^{2}q^{3}-(q^{1})^{2}q^{2}%
q^{3}\nonumber
\end{align}
with the corresponding canonical operator $\theta_{0}$ (\ref{3.1}) and
$\theta_{1}$ of the form
\[
\theta_{1}=\frac{1}{2}\left[
\begin{array}
[c]{cccccc}%
0 & 0 & 0 & q^{1} & -1 & 0\\
0 & 0 & 0 & q^{2} & 0 & -1\\
0 & 0 & 0 & 2q^{3} & q^{2} & q^{1}\\
-q^{1} & -q^{2} & -2q^{3} & 0 & p_{2} & p_{3}\\
1 & 0 & -q^{2} & -p_{2} & 0 & 0\\
0 & 1 & -q^{1} & -p_{3} & 0 & 0
\end{array}
\right]  .
\]
and with $\rho_{1}=-q^{1}$, $\rho_{2}=\tfrac{1}{4}(q^{1})^{2}+\tfrac{1}%
{2}q^{2}$, $\rho_{3}=-\tfrac{1}{4}q^{1}q^{2}-\tfrac{1}{4}q^{3}$.

From the form of $H_{1}$ one can directly see that the inverse metric tensor
$G$ expressed in $(q,p)$ variables has in this example an anti-diagonal form
\begin{equation}
G=\frac{1}{2}\left[
\begin{array}
[c]{ccc}%
0 & 0 & 1\\
0 & 1 & 0\\
1 & 0 & 0
\end{array}
\right]  \label{prz1Gqp}%
\end{equation}
while the conformal Killing tensor $L$ has the form
\[
L=\frac{1}{2}\left[
\begin{array}
[c]{ccc}%
q^{1} & -1 & 0\\
q^{2} & 0 & -1\\
2q^{3} & q^{2} & q^{1}%
\end{array}
\right]
\]
which substituted in (\ref{3.4a}) yields the geodesic parts of all the
Hamiltonians (\ref{hamex1}). Our quasi-bi-Hamiltonian system turns out to be
separable in $(\lambda,\mu)$ coordinates defined as above and it turns out
that in this case $f_{i}=1/8=const$ and $\gamma_{i}(\lambda^{i})=16(\lambda
^{i})^{7}$ in (\ref{pelnehamwlm}). Thus, we can easily find out the
constraints $\rho_{3}=v_{3}=0$ directly in $(q,p)$ coordinates. From the form
of $H_{3}$ in (\ref{hamex1}) one can see that $\varphi_{1}^{\prime\prime}%
=\rho_{3}(q)=-\frac{1}{4}(q_{1}q_{2}+q_{3})$ and that $\varphi_{2}%
^{\prime\prime}=v_{3}(q,p)$ is just the $g$-consequence of $\rho_{3}$. An easy
computation of (\ref{redM}) with the use of (\ref{prz1Gqp}) yields that
$v_{3}(q,p)=-\frac{1}{8}(p_{1}+p_{2}q^{1}+p_{3}q^{2})$. Performing the Dirac
deformation (\ref{2.10}) of $\theta_{0}$ and $\theta_{1}$ with the use of
constraints $\varphi_{1}^{\prime\prime}$ and $\varphi_{2}^{\prime\prime}$
yields
\[
\theta_{0,D}=\frac{1}{2q^{2}+(q^{1})^{2}}\left[
\begin{array}
[c]{c|c}%
\begin{array}
[c]{ccc}%
0 & 0 & 0\\
0 & 0 & 0\\
0 & 0 & 0
\end{array}
&
\begin{array}
[c]{ccc}%
q^{2}+(q^{1})^{2} & -q^{1} & -1\\
-q^{1}q^{2} & 2q^{2} & -q^{1}\\
-(q^{2})^{2} & -q^{1}q^{2} & q^{2}+(q^{1})^{2}%
\end{array}
\\\hline
-\ast^{T} &
\begin{array}
[c]{ccc}%
0 & q^{1}p_{2}-q^{2}p_{3} & p_{2}\\
q^{2}p_{3}-q^{1}p_{2} & 0 & p_{3}\\
-p_{2} & -p_{3} & 0
\end{array}
\end{array}
\right]
\]
and
\[
\theta_{1,D}=\frac{1}{2}\left[
\begin{array}
[c]{cccccc}%
0 & 0 & 0 & q^{1} & -1 & 0\\
0 & 0 & 0 & q^{2} & 0 & -1\\
0 & 0 & 0 & -2q^{1}q^{2} & q^{2} & q^{1}\\
-q^{1} & -q^{2} & 2q^{1}q^{2} & 0 & p_{2} & p_{3}\\
1 & 0 & -q^{2} & -p_{2} & 0 & 0\\
0 & 1 & -q^{1} & -p_{3} & 0 & 0
\end{array}
\right]
\]
so that $\theta_{1,D}$ differs from $\theta_{1}$ only at entries (3,4) and
(4,3), where $q^{3}$ was deformed to $-q^{1}q^{2}$ (notice that on
$\mathcal{S}$ indeed we have $q^{3}=-q^{1}q^{2}$). We will now pass to the
Casimir variables chosen as variables
\begin{equation}
(q^{1},q^{2},\varphi_{1}^{\prime\prime}(q),\varphi_{2}^{\prime\prime
}(q,p),p_{2},p_{3})\label{newvarex1}%
\end{equation}
since due to the fact that it is easiest to eliminate $q^{3}$ and $p_{1}$ from
the system of equations $\varphi_{1}^{\prime\prime}=\varphi_{1}^{\prime\prime
}(q),$ $\varphi_{2}^{\prime\prime}=\varphi_{2}^{\prime\prime}(q,p)$ \ we will
parametrize our submanifold by the coordinates $(q^{1},q^{2},p_{2},p_{3})$. In
the variables (\ref{newvarex1}) the operators $\theta_{i,D}$ attain the form
\[
\theta_{0,D}=\frac{1}{2q^{2}+(q^{1})^{2}}\left[
\begin{array}
[c]{cccccc}%
0 & 0 & 0 & 0 & -q^{1} & -1\\
0 & 0 & 0 & 0 & 2q^{2} & -q^{1}\\
0 & 0 & 0 & 0 & 0 & 0\\
0 & 0 & 0 & 0 & 0 & 0\\
q^{1} & 2q^{2} & 0 & 0 & 0 & p_{3}\\
1 & q^{1} & 0 & 0 & -p_{3} & 0
\end{array}
\right]
\]
and
\[
\theta_{1,D}=\frac{1}{2}\left[
\begin{array}
[c]{cccccc}%
0 & 0 & 0 & 0 & -1 & 0\\
0 & 0 & 0 & 0 & 0 & -1\\
0 & 0 & 0 & 0 & 0 & 0\\
0 & 0 & 0 & 0 & 0 & 0\\
1 & 0 & 0 & 0 & 0 & 0\\
0 & 1 & 0 & 0 & 0 & 0
\end{array}
\right]
\]
respectively. Observe that in these new variables all the entries at third and
fourth rows and columns are zero, as it should be, since now the Casimirs
$\varphi_{1}^{\prime\prime}$ and $\varphi_{2}^{\prime\prime}$ are part of our
coordinate system. We can now write down the reduced operators $\theta_{0,R}$
and $\theta_{1,R}$ in variables $(q^{1},q^{2},p_{2},p_{3})$ by simply removing
the zero rows and columns (we would also have to put $\varphi_{i}%
^{\prime\prime}=0$ but our matrices $\theta_{i,D}$ do not contain any
variables $\varphi_{i}^{\prime\prime}$ in their entries). This yields
\[
\theta_{0,R}=\frac{1}{2q^{2}+(q^{1})^{2}}\left[
\begin{array}
[c]{cccc}%
0 & 0 & -q^{1} & -1\\
0 & 0 & 2q^{2} & -q^{1}\\
q^{1} & -2q^{2} & 0 & p_{3}\\
1 & q^{1} & -p_{3} & 0
\end{array}
\right]  \text{ \ , \ }\theta_{1,R}=-\frac{1}{2}\left[
\begin{array}
[c]{cc}%
0_{2} & I_{2}\\
-I_{2} & 0_{2}%
\end{array}
\right]
\]
so that $\theta_{1,R}$ attains in the variables $(q^{1},q^{2},p_{2},p_{3})$
the canonical form. An easy calculation yields that the Hamiltonians $H_{i}$
restricted to $\mathcal{S}$ become
\begin{align*}
H_{1,R}(q^{1},q^{2},p_{2},p_{3}) &  =H_{1}|_{\mathcal{S}}=-q^{1}p_{2}%
p_{3}+\tfrac{1}{2}p_{2}^{2}-q^{2}p_{3}^{2}-20(q^{1})^{3}q^{2}\\
&  +12q^{1}(q^{2})^{2}+3(q^{1})^{5}\\
H_{2,R}(q^{1},q^{2},p_{2},p_{3}) &  =H_{2}|_{\mathcal{S}}=\left(  q^{2}%
+\tfrac{1}{2}(q^{1})^{2}\right)  p_{2}p_{3}+9(q^{1})^{2}(q^{2})^{2}\\
&  +\tfrac{5}{2}(q^{1})^{4}q^{2}-\tfrac{5}{4}(q^{1})^{6}-2(q^{2})^{3}\\
H_{3,R}(q^{1},q^{2},p_{2},p_{3}) &  =H_{3}|_{\mathcal{S}}=0
\end{align*}
while the functions $\rho_{i,R\text{ }}$are: $\rho_{1,R}=-q^{1}$, $\rho
_{2,R}=\tfrac{1}{4}(q^{1})^{2}+\tfrac{1}{2}q^{2}$, $\rho_{3,R}=0$. Thus, we
have obtained on $S$ a new separable quasi-bi-Hamiltonian system of the form
(\ref{qBHV}) with $n=2$. This concludes our first example.

As a second example we will consider a separable (quasi)-bi-Hamiltonian system
with a non-trivial function $f$. More specifically, we shall consider a
Lagrangian system of second order Newton equations
\begin{align*}
q_{,tt}^{1} &  =16q^{1}(1+q^{2})\\
q_{,tt}^{2} &  =8(q^{1})^{2}+4(q^{3})^{2}+64q^{2}+48(q^{2})^{2}+4c+24\\
q_{,tt}^{3} &  =4q^{3}(3+4q^{2})
\end{align*}
This system is a part of a quasi-bi-Hamiltonian chain with Hamiltonians
\begin{align}
H_{1} &  =p_{1}^{2}+p_{2}^{2}+p_{3}^{2}-4(q^{1})^{2}(1+q^{2})-(q^{3}%
)^{2}(3+4q^{2})\nonumber\\
&  -4\left(  3q^{2}+4(q^{2})^{2}+2(q^{2})^{3}\right) \nonumber\\
H_{2} &  =-\left(  2q^{2}+1\right)  p_{1}^{2}+2q^{1}p_{1}p_{2}-p_{2}%
^{2}+2q^{3}p_{2}p_{3}-2p_{3}^{2}(q^{2}+1)\label{hamwqpex2}\\
&  -(q^{1})^{2}\left(  (q^{1})^{2}+2(q^{3})^{2}+4(q^{2})^{2}+4q^{2}+2\right)
-(q^{3})^{2}\left(  (q^{3})^{2}+4(q^{2})^{2}+2q^{2}\right) \nonumber\\
&  +8(q^{2})^{3}+16(q^{2})^{2}+12q^{2}\nonumber\\
H_{3} &  =-(q^{3})^{2}p_{1}^{2}+2q^{1}q^{3}p_{1}p_{3}-2q^{3}p_{2}p_{3}+\left(
1+2q^{2}-(q^{1})^{2}\right)  p_{3}^{2}\nonumber\\
&  +\left(  (q^{1})^{2}+4(q^{2})^{2}+(q^{3})^{2}+6q^{2}+3\right)  (q^{3}%
)^{2},\nonumber
\end{align}
with the canonical Poisson tensors $\theta_{0}$ and%
\[
\theta_{1}=\left[
\begin{array}
[c]{cccccc}%
0 & 0 & 0 & 1 & q^{1} & 0\\
0 & 0 & 0 & q^{1} & 2q^{2}+1 & q^{3}\\
0 & 0 & 0 & 0 & q^{3} & 0\\
-1 & -q^{1} & 0 & 0 & -p_{1} & 0\\
-q^{1} & -2q^{2}-1 & -q^{3} & p_{1} & 0 & p_{3}\\
0 & -q^{3} & 0 & 0 & -p_{3} & 0
\end{array}
\right]
\]
and with the functions%
\[
\rho_{1}(q)=-2(q^{2}+1)\text{ \ },\text{ \ }\rho_{2}(q)=1-(q^{3})^{2}%
-(q^{1})^{2}+2q^{2}\text{ },\text{ }\rho_{3}(q)=(q^{3})^{2}%
\]
The above system is a particular example of so called bi-cofactor systems
\cite{PoissonJMP}, \cite{stefanjahans}, \cite{hans}, \cite{hans2},
\cite{Crampin} separated recently in \cite{zMaciejem}. In Darboux-Nijenhuis
coordinates $(\lambda,\mu)$ it turns out that now $f(\lambda^{i})=4\lambda
^{i}(\lambda^{i}-1)$ and $\gamma_{i}(\lambda^{i})=(\lambda^{i})^{5}$. This
means that in order to find the constraint $\varphi_{2}^{\prime\prime}(q,p)$
we have to use the formula (\ref{nuwqp}) with $\varphi_{1}^{\prime\prime
}(q)=(q^{3})^{2}$. Plugging these functions into (\ref{nuwqp}) and using the
above forms of $G$ and $F$ we get a rather complicated expression for
$\varphi_{2}^{\prime\prime}$%
\[
\varphi_{2}^{\prime\prime}(q,p)\equiv v_{3}(q,p)=\frac{1}{4}\left(  p_{1}%
\frac{(q^{1})^{2}+(q^{3})^{2}}{q^{1}}-p_{2}+p_{3}\left(  -q^{3}+\frac
{2q^{2}+(q^{1})^{2}-1}{q^{3}}\right)  \right)
\]
Notice that the constraint $\varphi_{2}^{\prime\prime}$ seems to have a
singularity on the surface $\varphi_{1}^{\prime\prime}=0$. However, it turns
out that on the submanifold $\mathcal{S}=\left\{  \varphi_{1}=\varphi
_{2}=0\right\}  $ we have $p_{3}=0$ and the singularity disappears (see
below). Having obtained the constraints $\varphi_{1}^{\prime\prime}$ and
$\varphi_{2}^{\prime\prime}$ in $(q,p)$ variables we now proceed as in the
first example. First we find
\[
\theta_{0,D}=\left[
\begin{array}
[c]{cccccc}%
0 & 0 & 0 & 1 & 0 & \zeta_{1}\\
0 & 0 & 0 & 0 & 1 & \zeta_{2}\\
0 & 0 & 0 & 0 & 0 & 0\\
-1 & 0 & 0 & 0 & 0 & \zeta_{3}\\
0 & -1 & 0 & 0 & 0 & \zeta_{4}\\
-\zeta_{1} & -\zeta_{2} & 0 & -\zeta_{3} & -\zeta_{4} & 0
\end{array}
\right]
\]
(where $\zeta_{i}$ are some rational functions of $q$ and $p$ that do not
vanish on $\mathcal{S}$) and
\[
\theta_{1,D}=\left[
\begin{array}
[c]{cccccc}%
0 & 0 & 0 & 1 & \frac{(q^{3})^{2}}{q^{1}}+q^{1} & 0\\
0 & 0 & 0 & q^{1} & 2q^{2}+1 & q^{3}\\
0 & 0 & 0 & 0 & 0 & 0\\
-1 & -q^{1} & 0 & 0 & p_{1}\left(  \frac{(q^{3})^{2}}{(q^{1})^{2}}-1\right)  &
0\\
-\frac{(q^{3})^{2}}{q^{1}}-q^{1} & -2q^{2}-1 & 0 & p_{1}\left(  1-\frac
{(q^{3})^{2}}{(q^{1})^{2}}\right)  & 0 & 2p_{3}\\
0 & -q^{3} & 0 & 0 & -2p_{3} & 0
\end{array}
\right]
\]
(notice that in both cases $q^{3}$ is a Casimir function). It can be checked
that the Schouten bracket of $\theta_{0,D}$ and $\theta_{1,D}$ is equal to
zero: $\left[  \theta_{0,D}\text{ },\text{ }\theta_{1,D}\right]  _{S}=0$ so
that the operators $\theta_{0,D}$ and $\theta_{1,D}$ are compatible. Before
projecting $\theta_{0,D}$ and $\theta_{1,D}$ onto $\mathcal{S}$ we perform a
transformation of variables to Casimir variables
\begin{equation}
(q^{1},q^{2},\varphi_{1}^{\prime\prime}(q),p_{1},p_{2},\varphi_{2}%
^{\prime\prime}(q,p))\label{newvarex2}%
\end{equation}
since this time it is easier to eliminate $q^{3}$ and $p_{3}$ from the system
of equations $\varphi_{1}^{\prime\prime}=\varphi_{1}^{\prime\prime}(q),$
$\varphi_{2}^{\prime\prime}=\varphi_{2}^{\prime\prime}(q,p)$. Explicitly we
have
\begin{align}
q^{3} &  =\pm\sqrt{\varphi_{1}^{\prime\prime}}\label{p3q3ex2}\\
p_{3} &  =\sqrt{\varphi_{1}^{\prime\prime}}\frac{4\varphi_{2}^{\prime\prime
}q^{1}+\varphi_{1}^{\prime\prime}p_{1}+(q^{1})^{2}p_{1}-q^{1}p_{2}}%
{2q^{1}q^{2}+q^{1}-q^{1}\varphi_{1}^{\prime\prime}-(q^{1})^{3}}\nonumber
\end{align}
As we have mentioned, we have now $p_{3}|_{\mathcal{S}}=0$. We will thus
parametrize our submanifold $\mathcal{S}$ by the coordinates $(q^{1}%
,q^{2},p_{1},p_{2})$. After expressing the operators $\theta_{i,D}$ in the
variables (\ref{newvarex2}) with the help of relations (\ref{p3q3ex2}) we can
easily project these operators on $\mathcal{S}$. As the result we get a new
(reduced) quasi-bi-Hamiltonian chain of the form (\ref{qBHV}) with \ $n=2$,
which is the variables $(q^{1},q^{2},p_{1},p_{2})$ is determined by
\[
\theta_{0,R}=\left[
\begin{array}
[c]{cc}%
0_{2} & I_{2}\\
I_{2} & 0_{2}%
\end{array}
\right]  \text{ \ , \ }\theta_{1,R}=\left[
\begin{array}
[c]{cccc}%
0 & 0 & 1 & q^{1}\\
0 & 0 & q^{1} & 2q^{2}+1\\
-1 & -q^{1} & 0 & -p_{1}\\
-q^{1} & -2q^{2}-1 & p_{1} & 0
\end{array}
\right]
\]
and by the restricted Hamiltonians $H_{i}$
\begin{align*}
H_{1,R} &  =H_{1}|_{\mathcal{S}}=p_{1}^{2}+p_{2}^{2}-4(q^{1})^{2}%
(1+q^{2})-4\left(  3q^{2}+4(q^{2})^{2}+2(q^{2})^{3}\right) \\
H_{2,R} &  =H_{2}|_{\mathcal{S}}=-(1+2q^{2})p_{1}^{2}+2q^{1}p_{1}p_{2}%
q-p_{2}^{2}\\
&  \;\;\;\;\;\;\;-(q^{1})^{2}\left(  (q^{1})^{2}+2(q^{3})^{2}+4(q^{2}%
)^{2}+4q^{2}+2\right) \\
&  \;\;\;\;\;\;\;+8(q^{2})^{3}+16(q^{2})^{2}+12q^{2}\\
H_{3,R} &  =H_{3}|_{\mathcal{S}}=0
\end{align*}
while the functions $\rho_{i,R}\,$ are given by $\rho_{1,R}=-2c(q^{2}+1)$,
$\rho_{2,R}=1+2q^{2}-(q^{1})^{2}$, $\rho_{3,R}=0$.

Notice that when we take $\varphi_{2}^{\prime\prime}(q,p)=2q^{3}p_{3},$ i.e.
the differential consequence of $\varphi_{1}^{\prime\prime}(q,p)=(q^{3})^{2},$
then $\{\varphi_{1}^{\prime\prime},\varphi_{2}^{\prime\prime}\}_{\theta_{1}%
}=0$ and the Dirac deformation $\theta_{1,D}$ does not exists (see Remark
\ref{dlaczego}).

\section{Conclusions}

In this article we have focused on the problem of Dirac deformation and Dirac
reduction of Poisson operators and Poisson pencils. We have presented a
procedure of performing Dirac reduction in quasi-bi-Hamiltonian systems of
Benenti type that do not destroy the separability of these systems and that
moreover do not change the separation variables. The method presented is not,
however, general in the sense that it does not provide us with all such
reductions, but only with their subclass. Moreover, for the moment the
procedure works only for those systems, for which the functions $f_{i}$ are
all equal, i.e. when $f_{i}$ does not depend on $i$. Last but not least, the
presented procedure is valid only in case when one of the Poisson structures
of the system is canonical - the non-canonical case must be studied
separately. These issues will be addressed in a separate paper.


\bigskip

\begin{thebibliography}{99}                                                                                               %
\bibitem {morositondo}C. Morosi, G. Tondo, ''Quasi-bi-Hamiltonian systems and
separability'', \emph{J. Phys. A: Math. Gen.} \textbf{30}, (1997), 2799.

\bibitem {1}B\l aszak M, ''On separability of bi-Hamiltonian chain with
degenerated Poisson structures'', \emph{J. Math. Phys.} \textbf{39, }3213 (1998)

\bibitem {2}B\l aszak M, ''Bi-Hamiltonian separable chains on Riemannian
manifolds'',\emph{\ Phys. Lett. A} \textbf{243, }25 (1998)

\bibitem {5}B\l aszak M, ''Theory of separability of multi-Hamiltonian
chains'',\emph{\ J. Math. Phys.} \textbf{40} (1999) 5725.

\bibitem {4}B\l aszak M, ''Inverse bi-Hamiltonian separable
chains'',\emph{\ J. Theor. Math. Phys.} \textbf{122 }(2000) 140.

\bibitem {6}B\l aszak M, ''Separability of two-Casimir bi- and tri-Hamiltonian
chains'',\emph{\thinspace} \emph{Rep. Math. Phys.} \textbf{46 }(2000) 35.

\bibitem {7}B\l aszak M, ''Degenerate Poisson Pencils on Curves: New
Separability Theory'', \emph{J. Nonl. Math.Phys.} \textbf{7 }(2000). 213

\bibitem {m1}Falqui G, Magri F. and Tondo G, ''Reduction of bihamiltonian
systems and separation of variables: an example from the Boussinesq
hierarchy'',\emph{\ } \emph{Theor. Math. Phys}. \textbf{122 }(2000) 176.

\bibitem {m2}Falqui G, Magri F, Pedroni M, ''Bihamiltonian geometry and
separation of variables for Toda lattices'',\emph{\ eprint} \ nlin.SI/0002008 (2000).

\bibitem {8}B\l aszak M, ''From bi-Hamiltonian geometry to separation of
variables: stationary Harry-Dym and the KdV dressing chain'',\emph{\ }%
\textbf{\ } \emph{Journal of Nonlinear Mathematical Physics,} \textbf{9
}Supplement \textbf{1 }\emph{\ }(2002) 1

\bibitem {ibort}Ibort A, Magri F., Marmo G, ''Bihamiltonian structures and
St\"{a}ckel separability'', \emph{Journal of Geometry and Physics} \textbf{33}
(2000) 210--228

\bibitem {statKdV}Falqui G., Magri F., Pedroni M., Zubelli J.P., ''A
Bi-Hamiltonian Theory for Stationary KdV flows and their Separability'',
\emph{Regular Chaotic Dynamics }\textbf{5 }(2000) 33

\bibitem {m3}Falqui G, Magri F, Pedroni M, ''Bihamiltonian geometry and
separation of variables for Toda lattices'',\emph{\ eprint} \ nlin.SI/0002008 (2000)

\bibitem {m4}Falqui G, Magri F. and Tondo G, ''Reduction of bihamiltonian
systems and separation of variables: an example from the Boussinesq
hierarchy'',\emph{\ } \emph{Theor. Math. Phys}. \textbf{122 }(2000) 176

\bibitem {zMaciejem}K. Marciniak, M. Blaszak, ''Separation of variables in
quasi-potential systems'', \emph{\ J. Phys. A: Math. Gen.} \textbf{35 }(2001) 2947

\bibitem {Dirac}P. A. M. Dirac, ''Generalized Hamiltonian Dynamics'',
\emph{Can. J. Math}. \textbf{2} (1950) 129--148

\bibitem {Dirac1}P. A. M. Dirac, Lecture Notes on Quantum Mechanics, Yeshova
University, New York, 1964

\bibitem {Magri}F. Magri, ''A simple model of the integrable Hamiltonian
equation'', \emph{J. Math. Phys.} \textbf{19} (5), 1978, pp. 1156--1162

\bibitem {MarsdenRatiu}J. Marsden, T. Ratiu, \textquotedblright Reduction of
Poisson manifolds\textquotedblright, $\emph{Lett.Math.}$\emph{$\pi$%
}$\emph{hys}$. \emph{11} (1986) 161--169

\bibitem {FalquiPedroni}G. Falqui, M. Pedroni, ''On Poisson reduction for
Gel'fand-Zakharevich manifolds'', to appear in \emph{Rep. Math. Phys.}

\bibitem {b1}S. Benenti, ''Orthogonal separable dynamical systems'', in:
Differential geometry and its applications, Proceedings on the Conference in
Opava, 24-28 August 1992, Silesian University, Opava, vol. I, 1993, p. 163;
Electronic edition http:/www.emis.de/proceedings/

\bibitem {b2}S. Benenti, ''Intrinsic characterization of the variable
separation in the Hamilton-Jacobi equation'', \emph{J. Math. Phys. }\textbf{39
}(1997) 6578-6602

\bibitem {hans}H. Lundmark, ''A new class of integrable Newton systems'',
\emph{Journal of Nonlinear Mathematical Physics}, \textbf{8}, (2001) 195-199,
Supplement - Proceedings of NEEDS '99, Kolymbari, Crete.

\bibitem {HF1}P. Vanhaecke, ''Integrable systems in the Realm of Algebraic
Geometry'', Lecture Notes in Mathematics 1638, Springer-Verlag, New York, 1996

\bibitem {HF2}F. Magri and T. Marsico, ''Some Developments of the Concept of
Poisson Manifolds in the Sense of A. Lichnerowicz'', in Gravitation,
Electromagnetism and Geometric Structures, ed. G. Ferrarese, Pitagora
editrice, Bologna, 1996, p.207

\bibitem {myPhysicaA}S. Rauch-Wojciechowski, K. Marciniak, M. Blaszak, ''Two
Newton decompositions of stationary flows of KdV and Harry Dym
hierarchies''\ \emph{Physica A} \textbf{233} (1996) pp.307--330

\bibitem {PoissonJMP}Marciniak K, Rauch-Wojciechowski S, ''Two families of
nonstandard Poisson structures for Newton equations'', \emph{J. Math.
Phys}.\textbf{\ 39} (10), 1998, 5292--5306

\bibitem {stefanjahans}Rauch-Wojciechowski S, Marciniak K, Lundmark H,
''Quasi-Lagrangian systems of Newton equations'', \emph{J. Math. Phys}.
\textbf{40} (12), 1999, 6366--6398.

\bibitem {hans2}H. Lundmark, ''Higher-dimensional integrable Newton systems
with quadratic integrals of motion'', to appear in $\emph{Stud.Appl.Math}$.

\bibitem {Crampin}M. Crampin, W. Sarlet, ''A class of non-conservative
Lagrangian systems on Riemannian manifolds'', J. Math. Phys. 42 (9), 2001, pp. 4313--4326.
\end{thebibliography}
\end{document}